\documentclass[draft]{agujournal2019}
\usepackage{url} 
\usepackage{lineno}
\usepackage{amsmath}
\usepackage[inline]{trackchanges} 
\usepackage{soul}
\draftfalse

\journalname{Journal of Advances in Modeling Earth Systems (JAMES)}

\begin{document}

%
%


\title{Stable Machine-Learning Parameterization of Subgrid Processes in a Comprehensive Atmospheric Model Learned From Embedded Convection-Permitting Simulations}

%
%




\authors{Zeyuan Hu\affil{1,2}\thanks{Work done during an internship at NVIDIA}, Akshay Subramaniam\affil{1}, Zhiming Kuang\affil{2}, Jerry Lin\affil{3}, Sungduk Yu\affil{4}, Walter M. Hannah\affil{5}, Noah D. Brenowitz\affil{1}, Josh Romero\affil{1}, Michael S. Pritchard\affil{1,3}}

\affiliation{1}{NVIDIA Research}
\affiliation{2}{Harvard University}
\affiliation{3}{University of California at Irvine}
\affiliation{4}{Multimodal Cognitive AI, Intel Labs, Santa Clara, CA 95054, USA}
\affiliation{5}{Lawrence Livermore National Laboratory}





\correspondingauthor{Zeyuan Hu}{zeyuan\_hu@fas.harvard.edu}




\begin{keypoints}
\item Stable hybrid climate simulations are achieved with a data-driven emulator of subgrid physics coupled with a comprehensive atmosphere model.
\item Online performance benefits from a U-Net architecture and microphysical constraints.
\item A realistic cloud climatology with explicit cloud condensate coupling is achieved in a hybrid multi-scale modeling framework.
\end{keypoints}

%
%

%
%


\begin{abstract}
Modern climate projections often suffer from inadequate spatial and temporal resolution due to computational limitations, resulting in inaccurate representations of sub-grid processes. A promising technique to address this is the Multiscale Modeling Framework (MMF), which embeds a kilometer-resolution cloud-resolving model within each atmospheric column of a host climate model to replace traditional convection and cloud parameterizations. Machine learning (ML) offers a unique opportunity to make MMF more accessible by emulating the embedded cloud-resolving model and reducing its substantial computational cost. Although many studies have demonstrated proof-of-concept success of achieving stable hybrid simulations, it remains a challenge to achieve near operational-level success with real geography and comprehensive variable emulation that includes, for example, explicit cloud condensate coupling. In this study, we present a stable hybrid model capable of integrating for at least 5 years with near operational-level complexity, including coarse-grid geography, seasonality, explicit cloud condensate and wind predictions, and land coupling. Our model demonstrates skillful online performance, achieving a 5-year zonal mean tropospheric temperature bias within 2K, water vapor bias within 1 g/kg, and a precipitation RMSE of 0.96 mm/day. Key factors contributing to our online performance include an expressive U-Net architecture and physical thermodynamic constraints for microphysics. With microphysical constraints mitigating unrealistic cloud formation, our work is the first to demonstrate realistic multi-year cloud condensate climatology under the MMF framework. Despite these advances, online diagnostics reveal persistent biases in certain regions, highlighting the need for innovative strategies to further optimize online performance.
\end{abstract}

\section*{Plain Language Summary}

Traditional climate models often struggle to accurately simulate small-scale processes like thunderstorms due to compute limitations, leading to less reliable climate predictions. Machine learning (ML) offers a promising solution by efficiently modeling these processes and integrating them into hybrid ML-physics simulations within a host climate model. While previous studies have shown success in simplified setups, such as all-ocean planets, achieving accurate and stable hybrid simulations in complex, real-world settings remains challenging. In this study, we developed a stable hybrid model capable of simulating the climate for five years using real geographic features and explicitly predicting the time evolution of temperature, moisture, cloud, and wind. Our model achieves skillful accuracy in long-term mean atmospheric states. This success is due to several key improvements: an advanced architecture and the incorporation of cloud physics constraints.

%
%

%


%
%
%
%

\section{Introduction}

Accurately representing cloud and convection has long been a major challenge for accurate climate model simulations due to the relatively coarse resolution ($>$50km) in traditional climate models \cite{eyring2016overview}. Uncertainty associated with representing these subgrid processes remains one of the most significant factors contributing to climate feedback uncertainties in future climate change projections \cite{IPCC2021,Sherwood2020}. To better represent these processes, the community
has developed global storm-resolving simulators with $<$5km resolution. These models are
actively being tested for their use in long-term climate simulations, but they are still computationally too
demanding \cite<e.g.,>{taylor2023simple,hohenegger2023icon}. 

Another approach, known as the Multiscale Modeling Framework (MMF)—also commonly referred to as super-parameterization—explicitly embeds a small-domain, km-resolution cloud-resolving model (CRM) in each atmosphere column of a host climate model to replace the traditional convection and cloud parameterization \cite{Grabowski1999, Benedict2009,Randall2013,hannah2020initial,Norman2022}. The embedded cloud-resolving model can directly simulate the atmospheric processes at the km-scale with time steps of a few seconds and feedback to the host climate model the effect of these small-scale processes on the coarser grids. However, such MMF models can still be computationally very demanding, for example, 100-10,000 times more expensive than conventional global climate models depending on the dimensionality and resolution of the interior resolved scale \cite<e.g.,>{khairoutdinov2001cloud,parishani2018insensitivity,terai2020impact,hannah2020initial,peng2024improving}. To enhance computational efficiency, the embedded cloud-resolving model is often configured with a two-dimensional domain.

A hybrid approach that combines traditional climate models with machine learning is a promising solution for high-fidelity climate simulations without incurring additional computational costs. Machine learning models can learn the effects of small-scale physical processes directly from high-fidelity simulation data, such as those from MMF climate simulations \cite{gentine2018could,rasp2018deep,brenowitz2020interpreting,han2020moist,mooers2021assessing,wang2022stable,bhouri2023multi,lin2023sampling,iglesias2024causally,kuhbacher2024towards}, globally uniform km-resolution simulations \cite{brenowitz2018prognostic,brenowitz2019spatially,yuval2020stable,yuval2021use,wang2022non}, and reanalysis products \cite{kochkov2024neural}. Once trained, an ML emulator can be coupled with a host climate model to perform hybrid physics-ML simulations, potentially enhancing simulation accuracy for a given computational budget when comparing to traditional climate simulations using empirical convection parameterizations \cite{rasp2018deep,yuval2020stable}. 

The hybrid approach of learning a ML paramterization from MMF simulations has its own advantages and limitations compared to other hybrid methods. The MMF simulations might be considered less physical compared to global storm-resolving simulations due to certain idealizations, such as the scale separation between host model and the embedded CRM, and the 2D representation of clouds and convection in the 2D CRM. However, the MMF approach effectively captures the essence of the challenge of learning a ML parameterization, e.g, stochasticity, multi-physics, and coarse-graining in time. The scale separation also brings the convenience of locality, as CRMs in different columns do not directly communicate. Therefore, each CRM can be treated as an independent sample, and emulating each CRM would only require the input and output in the same column. 

Conversely, using global km-resolution simulation data requires coarse-graining the raw data, and the choice of coarse-graining scale can be challenging, potentially leading to issues when the ML parameterization learned from the coarse-graining data is coupled with a host model \cite<e.g,>{ross2023benchmarking}. The defined machine learning problem of emulating coarse-graining targets may also involve non-locality \cite{wang2022non}, where the state tendencies of each column can be dependent on the states of adjacent columns. In addition to learning from simulation data, \citeA{kochkov2024neural} trains their hybrid models, called Neural General Circulation Model (NeuralGCM), directly from reanalysis data. They achieved this by building a fully differentiable model, which includes a differentiable dynamical core and a neural network-based parameterization for physics in each column. The NeuralGCM has the advantage of optimizing climate metrics directly against reanalysis data of historical climate but faces challenges of extrapolating to substantially different future climate. In contrast, the MMF hybrid approach has a code base that is fit for climate prediction, with all necessary components and modules for simulating future climate beyond the historical record.

In addition to advantages and limitations mentioned in the previous paragraph, these hybrid physics-ML climate simulations face challenges related to online instability and online error. Optimizing a ML model's offline skill (by optimizing some loss function of the ML model's prediction at one single time step) does not necessarily lead to optimized online performance when a ML model is coupled with a dynamical core and integrates for years \cite{ott2020,wang2022stable}. Many previous studies have demonstrated proof-of-concept success in achieving stable online simulations using idealized setups such as aquaplanets. A few recent works \cite{wang2022stable,han2023ensemble,sanford2023improving,kochkov2024neural,behrens2024improving} have demonstrated the promise of stable online simulations incorporating real geographic features. \citeA{kochkov2024neural} achieved mostly stable (though occasional instability was observed) hybrid simulations using end-to-end online training, developing a fully differentiable hybrid model and directly optimizing multi-step rollout loss. Nevertheless, making existing Fortran-based climate models differentiable is challenging. 

In the context of emulating the CRM in MMF simulations with a ML parameterization, no previous work has succeeded in completely coupling all input and output variables of the CRM (e.g., tendencies of cloud condensates, surface fluxes for atmosphere-surface coupling) with real geography via ML. For example, \citeA{behrens2024improving} had to sidestep the condensate coupling to maintain a 5-month online stability by using a partial coupling approach, where condensate prediction is provided by a simultaneously running CRM. \citeA{han2023ensemble} predicted cloud species and achieved stable online simulations with real geography, but they did not emulate radiation (and thus the radiative fluxes required for atmosphere-land coupling) and showed significant online biases in high-latitude temperature and humidity. Additionally, \citeA{han2023ensemble} predicted cloud species in a semi-implicit manner, diagnosing current time step cloud states based on current temperature and moisture states (and many other input features) without interacting with the dynamical core, as indicated in Table 1 of \citeA{han2020moist}. Conventional cloud parameterizations were also retained in \citeA{han2023ensemble} to supply inputs required by their radiative transfer scheme—such as cloud liquid and ice number concentrations and cloud fraction—that were not predicted by their neural network.

In the long term, it is crucial for ML parameterizations to explicitly and accurately predict cloud condensate tendencies and radiative transfer, which are vital for handling aerosol cloud interaction, pollutant transport, and climate change. A recent open-source dataset, ClimSim \cite{yu2024climsim}, provides a valuable opportunity to train ML parameterizations with operational-level complexity. This dataset represents the largest-ever open-source collection tailored for operation-ready hybrid physics-ML climate simulations. It provides a comprehensive set of inputs and outputs for downstream coupling and contains an extensive number of samples with high sampling frequency at multiple spatial resolutions. It includes a coarse-resolution version (11.5°$\times$11.5°, 384 grid columns, $<1$ TB), suitable for cost-effective prototyping, and a high-resolution version (1.5°$\times$1.5°, 21,600 grid columns, 42 TB), which enables evaluation of performance under operational-level conditions with more detailed orography and land-sea contrasts.

In this work, we used the coarse-resolution version of the ClimSim dataset to train ML models capable of predicting a complete set of output variables. Our hybrid physics-ML simulations are stable under near-operational-level complexity (with a real but coarse geography, seasonality, explicit prediction of cloud condensate and wind tendencies, and surface coupling), outperforming previous studies that emulate the CRM in MMF climate simulations. Our 5-year stable hybrid simulation demonstrates less than 2K zonal mean troposphere temperature bias, less than 1g/kg zonal mean troposphere humidity bias, a root mean square error of 0.96 mm/day for mean precipitation, and realistic zonal mean cloud distribution. To inform others' practice, we show how this online performance is achieved as the result of several key efforts: employing an expressive U-Net architecture and incorporating cloud microphysical constraints.

The rest of the manuscript is structured as follows: Section 2 documents the methods, including details on model architecture, microphysical constraints, data, and training. Section 3 presents the offline skill improvement from using the more expressive U-Net architecture and expanded input features compared to a baseline multi-layer perceptron (MLP) model, a fully connected neural network architecture. Section 4 demonstrates the online performance of our hybrid simulations. Finally, Section 5 provides the discussion and summary.

\section{Methods}

\subsection{Dataset}
 
The data we used is the low-resolution real-geography dataset in the ClimSim dataset \cite{yu2024climsim}. This low-resolution real-geography data consists of 10-year simulation from the Energy Exascale Earth System Model (E3SM)-multiscale modeling framework (MMF) \cite{hannah2020initial}, configured with the predefined "F2010-MMF1" compset with ``ne4pg2'' grid and a model time step of 20 minutes. This low-resolution simulation contains 384 columns in the globe using an unstructured cubed-sphere grid, with an effective resolution of approximately 11.5°$\times$11.5°. The E3SM-MMF model employs 60 levels for global dynamics, extending up to approximately 65 km in altitude. The E3SM-MMF simulation uses prescribed sea surface temperature, sea ice coverage, and aerosols based on present-day climatology. The prescribed sea surface temperature follows a fixed recurring annual cycle. The aerosols are configured to be transparent to radiation code. The first seven years are used for the training set without subsampling. The eighth year is used for the validation set, while the last two years are used for the test set. Both the validation and test sets (not the training set) are subsampled in time using a stride of 7 time steps to be consistent with the constructed validation and test sets in \citeA{yu2024climsim}. Each simulation day has 72 time steps, so the stride of 7 adequately samples the full diurnal cycle. This whole low-resolution dataset provides around 100 million samples with a total data size of 744GB, of which 521GB is used for the training set. 

\subsection{Architecture}

\begin{figure}
\includegraphics[width=1\textwidth]{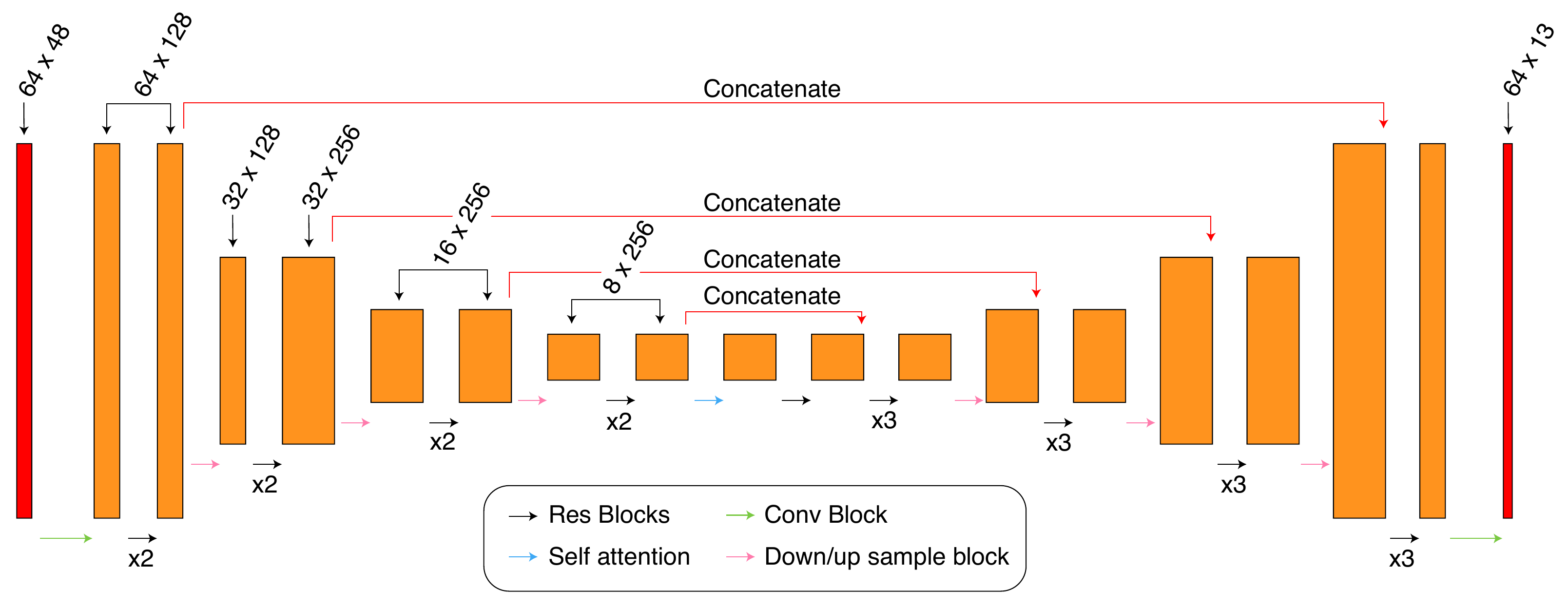}
\centering
\caption{
The schematic of the U-Net architecture. The U-Net model is composed of multiple ResBlocks, with 4 layers in depth and latent dimension sizes of [128, 256, 256, 256]. The model requires the input and output of each sample to be prepared as 2D arrays (see Appendix \ref{appendix:additional_preprocessing}) with dimensions corresponding to the number of variables times the vertical length. The numbers on top of each tensor indicate the tensor shape for the U-Net model with the microphysical constraint. Each tensor in the encoder has a skip connection to a corresponding tensor in the decoder, illustrated by the red arrows showing some of these skip connections.
}
\label{U-Net}
\end{figure}

Two architectures were used and compared in this study: a multi-layer perceptron (MLP) model and a U-Net model. The first architecture is the MLP model, which is a fully connected neural network with one or more hidden layers. We adopted the hyperparameters recommended by \cite{yu2024climsim}, which were informed by their ambitious hyperparameter search comprising 11,851 trials. MLP models are widely used in previous related studies \cite{gentine2018could,rasp2018deep,brenowitz2019spatially,ott2020,yuval2021use,beucler2024climate,lin2023sampling,iglesias2024causally} and serve well as a baseline for comparison. 

The second architecture is a U-Net model \cite{song2020score}, suitable for tasks where the output targets share the same spatial dimensions as the input features (here, the number of vertical levels in each atmospheric column). Other architectures have also been used in previous studies, such as Residual Convolutional Neural Networks (ResCu) in \citeA{han2020moist,han2023ensemble} and Residual Deep Neural Networks (ResDNNs) in \citeA{wang2022stable,han2023ensemble}. \citeA{heuer2023interpretable} recommended the U-Net architecture after comparing it with MLPs, Convolutional Neural Networks, and Residual Neural Networks, demonstrating that U-Net achieved the highest fitting score. We adapted the 2D U-Net model from \citeA{song2020score} into a 1D version (Figure \ref{U-Net}) by replacing 2D convolution layers with 1D convolution layers. The U-Net model has an encoder-decoder structure, where the encoder compresses the input data and the decoder reconstructs it, with skip connections that retain important details from the input. This setup enables the model to capture both short-range details, such as information from adjacent vertical levels, and long-range patterns, such as vertical structure of the entire column. Compared to the MLP model, the U-Net model can better learn the vertical structures through its convolutional layers and optional self-attention layers. Full details of the model parameters for both the MLP and U-Net models are documented in Appendix \ref{appendix:architecture}.

\subsection{CRM Setup in the E3SM-MMF Model}

Here we provide a brief description of the multi-scale modeling framework setup in the E3SM-MMF to help clarify the underlying emulation targets for machine learning models.

The E3SM-MMF climate model embeds a cloud-resolving model (CRM) in each atmospheric model column to replace parameterizations of microphysics, convection, radiation and turbulence. The embedded CRM is the System for Atmosphere Modeling model \cite<SAM, >{khairoutdinov2003cloud}. The CRM model is configured with a 2D domain of 64 columns and 1km horizontal resolution. While the E3SM has 60 vertical levels, the CRM only operates on the lowest 50 levels, excluding the uppermost 10 levels to circumvent issues associated with applying the anelastic approximation in areas of extremely low density. Therefore, the CRM output has zero tendencies for water species and momentum in the top 10 E3SM levels, but it still has non-zero tendency for temperature since the radiation calculation is conducted through the whole 60 levels. 

The CRM model uses a one-moment microphysics scheme \cite{khairoutdinov2003cloud}. The only prognostic water species in this one-moment scheme are the total water (vapor plus non-precipitating condensate) and total precipitating water. The one-moment scheme uses an instantaneous saturation adjustment to generate and remove cloud condensate on the internal CRM grid. Cloud condensate is then partitioned into liquid, ice, or mixed phase based on a diagnostic function of temperature as follows. Between 273.16K and 253.16K, partitioning of cloud condensate into cloud ice and liquid water depends linearly on temperature (at 253.16K, all condensate is ice; at 273.16K, all condensate is liquid water). Precipitating water is partitioned into rain, snow, and graupel in the same way. It is worth to note that the E3SM grid only tracks water vapor, cloud liquid, and cloud ice and does not track precipitating water.

The CRM in E3SM-MMF uses the Explicit Scalar Momentum Transport (ESMT) method \cite{tulich2015strategy} to represent convective momentum transport (CMT) for large-scale winds. ESMT treats the zonal and meridional components of large-scale momentum as non-conserved scalars, which are advected by resolved fine-scale flows within the 2D CRM. Since the 2D CRM lacks a third spatial dimension, the ESMT scheme implicitly assumes that the advective flow field has identical spatial structures in the zonal and meridional directions. Despite this limitation, \citeA{yang2022convective} demonstrated that the E3SM-MMF with a 2D CRM and ESMT produces a reasonable global distribution of CMT, comparable to simulations using a 3D CRM with fully explicit convective momentum transport. While the qualitative patterns of CMT tendencies are similar between the 2D and 3D approaches, the ESMT-derived tendencies are consistently weaker in magnitude compared to those produced by the 3D CRM.

In E3SM-MMF, radiation transfer calculations are performed at every GCM timestep (20 minutes) after the CRM module, separately on individual sub-columns of the CRM state. The calculated radiative heating tendencies are applied as constant forcings to the next timestep's CRM calculation. The radiation scheme from the GCM is used, taking the time-averaged state of each CRM column from the previous CRM calculation as input \cite{hannah2020initial}.

The CRM module replaces several parameterizations in conventional GCMs, including those for convection, cloud fraction, and microphysics. The CRM module also calculates the surface precipitation fluxes, which are passed to the land component. Our NN emulators aim to replace the entire CRM including its radiation calculation. 

Exterior to the CRM, on large-scales, the host atmosphere model in the E3SM-MMF retains conventional treatments of advection physics within the dynamical core, as well as additional physics modules that handle vertical diffusion and gravity wave drag. The dynamical core uses a spectral element method on a cubed-sphere geometry \cite{hannah2020initial}, updating all state variables including temperature, winds, water vapor, cloud liquid, and cloud ice.

\subsection{Microphysical Constraints}

\subsubsection{Liquid-ice Cloud Partition}

\begin{figure}
\includegraphics[width=1\textwidth]{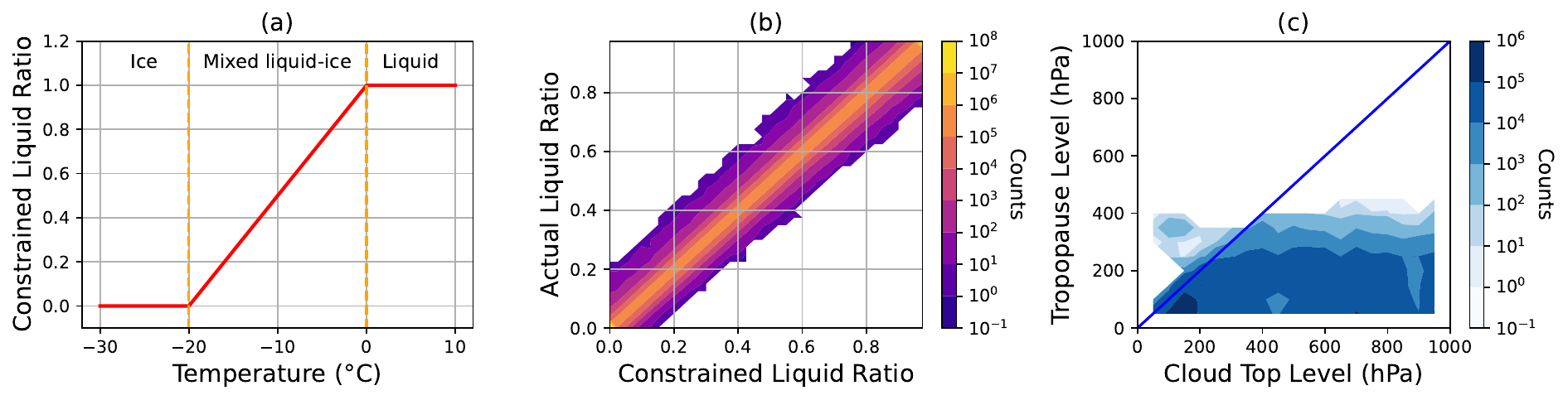}
\centering
\caption{
Assessment of the microphysical constraints. (a) The fraction of liquid over total cloud water mixing ratios as a function of temperature. This temperature-based phase partitioning holds exactly on each grid in the cloud-resolving model. (b) Histogram of the actual liquid cloud fraction in the E3SM grid (y-axis) versus the predicted liquid fraction based on the E3SM-grid temperature (x-axis). (c) Histogram of the tropopause level versus the cloud top level, where levels correspond to the hybrid pressure level used in E3SM. In each column, the tropopause level is identified as the lowest level where the pressure is less than 400 hPa and the vertical gradient of potential temperature exceeds 10 K/km. A cloud top is defined as the highest level with a total cloud water mixing ratio greater than 1 mg/kg.
}
\label{liq-partition}
\end{figure}

It is tempting to think of the emulation problem from the perspective of the host model, focusing on what the Global Climate Model (GCM) typically receives from the CRM. The ClimSim dataset \cite{yu2024climsim} was designed with this view in mind, providing the necessary data for seamless integration. However, our work demonstrates that it is especially important to consider the CRM and its own fundamental prognostics. In particular, the one-moment microphysics scheme in the CRM only predicts the total cloud (ice plus liquid) mixing ratio and uses temperature to diagnose the mixing ratios of liquid cloud and ice cloud on each CRM grid. According to the thermodynamic principles and CRM model code design, only liquid cloud is allowed to exist when the temperature is above 273.16K, and only ice cloud is allowed when the temperature is below 253.16K (Figure \ref{liq-partition}). In between, the fraction of liquid cloud over total cloud mixing ratio follows a linear function of temperature. 

While this temperature-based partition relationship between liquid and ice clouds is strictly enforced in every CRM grid, Figure \ref{liq-partition}b shows that this relationship also holds on the E3SM grids which reflect the horizontally averaged values in the CRM domain. The exact fraction of liquid cloud over total cloud condensate on each E3SM grid is well predicted by the temperature-based function, as most of the samples fall into either the lower left corner (pure ice cloud), the upper right corner (pure liquid cloud), or the diagonal (mixed-phase cloud). 

To leverage this microphysical constraint, instead of predicting liquid and ice separately, as the host GCM is accustomed to receiving its information from its embedded CRM in an MMF, we can predict only the change of total (i.e., liquid plus ice) cloud condensate at each step, a more fundamental total water prognostic of the CRM itself. After the neural network updates the temperature and total cloud mixing ratio in each grid, the updated temperature is then used to decompose the total into the amounts of liquid and ice clouds. This separation is necessary because the dynamical core advects liquid and ice clouds independently. This approach ensures that the correct cloud phase appears at the appropriate temperature range.

\subsubsection{Removing clouds above stable tropopause level}

In the hybrid online simulations, we implement a constraint of removing all the cloud above the tropopause to improve simulation stability. The tropopause is identified as the lowest level where the pressure is less than 400 hPa and the vertical gradient of potential temperature exceeds 10 K/km. The intuition of this constraint is that the tropopause acts as a stable barrier preventing deep convection from penetrating and forming clouds above it. Any clouds above the tropopause, whether formed by deep penetrating convection in previous time steps or advected from lower levels, tend to be removed by sedimentation processes. The empirically chosen thresholds for identifying the tropopause ensure that the tropopause generally lies above the cloud top in most cases (Figure \ref{liq-partition}c). The cloud top is defined as the highest level with a total cloud mixing ratio greater than 1 mg/kg. Most residual clouds above the defined tropopause after invoking the CRM have mixing ratios on the order of 0.1 mg/kg or smaller, while the zonal mean cloud mixing ratio in the upper troposphere is at the order of 10 mg/kg (see Section \ref{section:online}). This constraint does not alter the model architecture or training process but is implemented during online inference to enhance stability. Implementing this constraint allowed us to easily test its impact and test different thresholds for the defined tropopause without needing to retrain the model. However, it is also possible to incorporate this constraint directly into the ML model itself and both methods are equivalent.

\subsection{Inputs and Outputs}

The input features for the MLP model include GCM-scale state variables (e.g., temperature, relative humidity, cloud liquid mixing ratio, cloud ice mixing ratio, zonal wind, and meridional wind), atmospheric gas components, and boundary conditions (e.g., solar insolation, surface heat fluxes, surface stress, albedo, and geographic information). These GCM state variables are saved immediately prior to a code region containing the CRM and radiation physics. The output targets include the CRM's time tendencies of temperature, specific humidity, cloud liquid and ice mixing ratios, zonal and meridional winds, as well as radiative fluxes and surface precipitation rates. That is, the ML targets encompass all the forcings that the CRM transmits to its host GCM in the MMF approach to climate simulation, including challenging condensate tendencies.

\subsubsection{Expanding Inputs to Include Large-scale Forcings and Convection Memory}
The U-Net model uses expanded input features beyond ClimSim's standard input that also include large-scale forcings (primarily the advective tendencies of temperature, total water, and zonal wind from the dynamical core), convection memory (CRM tendencies from previous steps), and latitudes. 

\textbf{Large-scale forcings} refer to the tendencies of temperature, total water (water vapor plus liquid and ice cloud condensates), and zonal wind outside of the CRM in each time step, primarily from the dynamical core. During the CRM calculation, these large-scale forcings are applied uniformly in time and at each vertical level in the CRM domains. Such large-scale forcings can influence convection; for example, a large-scale forcing that stabilizes the atmosphere could make the convection more difficult to trigger, making them causally relevant input features. However, most previous studies of hybrid MMF simulations failed to include these features, with only a few exceptions \cite{han2020moist,wang2022stable,han2023ensemble}. 

\textbf{Convection memory} refers to the internal convection state within the CRM that is not adequately represented in the GCM grid state (which is the averaged value of the CRM domain). The internal convection state, including aspects such as internal cloud structure and convection organization, can be useful predictors for convective tendencies and precipitation \cite{shamekh2023implicit,kuang2024linear}. The CRM tendencies at previous time steps have been used in a few studies \cite{han2020moist,han2023ensemble,lin2023sampling} as a proxy to represent aspects of convection memory. 

Due to the physical relevance, we expect that adding these expanded input features can help with both offline and online skill. Both inputs and outputs are normalized to be on the order of 1. Detailed descriptions of the full list of input and output features and their normalization methods can be found in the Appendix (Sections \ref{appendix:inpu_output} and \ref{appendix:normalization}).

\subsection{Training}

We used and compared four sets of offline model designs: a baseline MLP architecture without expanded input features such as large-scale forcings or convection memory, a U-Net architecture without expanded input features (hereafter U-Net-baseline), a U-Net architecture with expanded input features (hereafter U-Net-expanded), and a U-Net architecture with both expanded input features and the cloud microphysics constraint of the temperature-based liquid-ice partition (hereafter U-Net-expanded-constrained). 

To account for stochasticity in ML optimization, we used three different checkpoints for each model architecture. These checkpoints were obtained using varying loss functions (either mean-absolute-error loss or Huber loss) and learning rate schedules, creating a small ensemble to account for some of the uncertainty in online performance stemming from offline ML optimization variability. All the models were trained in PyTorch \cite{paszke2019pytorch} using the NVIDIA Modulus framework \cite{modulus}. Please see \ref{appendix:training} for the detailed training hyperparameters for these different checkpoints. In the subsequent sections and plots where only one of the three checkpoints is shown, it is the one that uses the Huber loss and a reduce-on-plateau learning rate scheduler. Choosing other checkpoints does not change the following results qualitatively.

\subsection{Coupling E3SM and NN}

The trained PyTorch neural networks are integrated with the Fortran-based E3SM codes using an open-source library called Pytorch-Fortran \cite{torchfort}. This library requires saving the PyTorch model in TorchScript \cite{torchscript} serialized format, allowing the Fortran-based E3SM to directly read the TorchScript model and perform efficient batch inference within the Fortran process using bindings to the TorchScript C++ API. This coupling approach is compatible with most machine learning models written in PyTorch, allowing for extensive testing of diverse ML models. Furthermore, this E3SM-NN coupling workflow is integrated into a containerized, end-to-end framework called ClimSim-Online \cite{yu2024climsimonline} which allows researchers to run and test such hybrid E3SM simulations on multiple platforms.

\section{Offline results}

\begin{figure}
\includegraphics[width=1\textwidth]{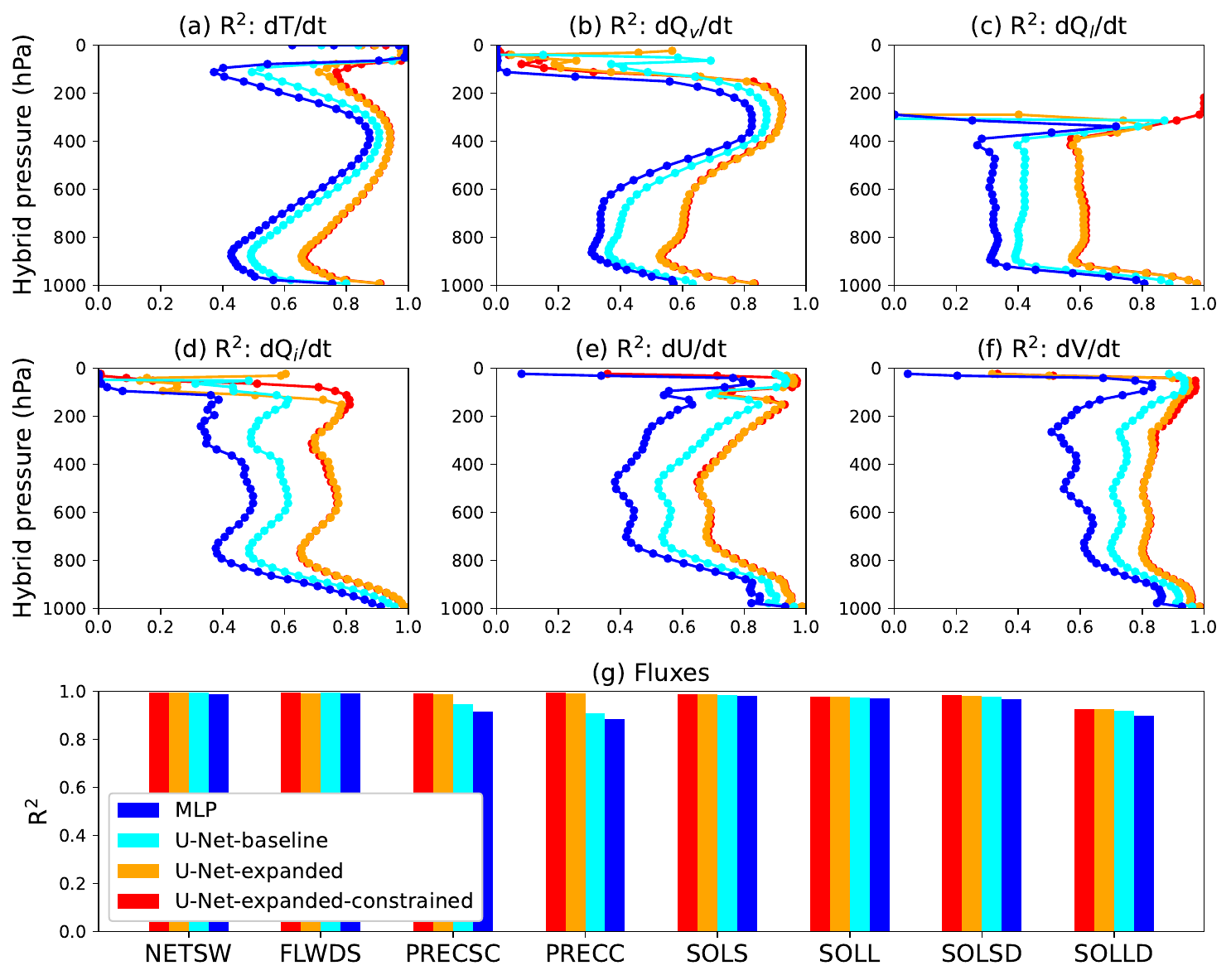}
\centering
\caption{
Offline R$^2$ scores across various variables for MLP (blue), U-Net-baseline (cyan), U-Net-expanded (orange), and U-Net-expanded-constrained (red). Variables are the full target variables listed in Table \ref{output_table}, including temperature tendency ($\frac{dT}{dt}$), water vapor tendency ($\frac{dQ_{\text{v}}}{dt}$), liquid cloud mixing ratio tendency ($\frac{dQ_{\text{l}}}{dt}$), ice cloud mixing ratio tendency ($\frac{dQ_{\text{i}}}{dt}$), zonal wind tendency ($\frac{dU}{dt}$), meridional wind tendency ($\frac{dV}{dt}$), and eight flux variables.
}
\label{r2-offline}
\end{figure}


Figure \ref{r2-offline} shows the comparison of R$^2$ across all target variables for the MLP and U-Net models. R$^2$ for each target variable at each vertical level is calculated using data across all the columns and time steps in the test set. For the U-Net-expanded-constrained model, the R$^2$ values reflect only the constraint involving liquid-ice cloud partitioning, as the constraint for removing stratospheric clouds is applied solely in online simulations. These R$^2$ results are from one of the checkpoints using the Huber loss function. 

Our baseline MLP model shows similar R$^2$ of temperature and lower tropospheric water vapor tendencies compared to the MLP models in \citeA{yu2024climsim}. However, our MLP model demonstrates significantly better R$^2$ of cloud condensates and wind tendencies. Figure 8 in the Supplementary Information in \citeA{yu2024climsim} shows non-positive R$^2$ at most levels for winds and cloud ice tendencies, as opposed to our MLP results in Figure \ref{r2-offline}. 

This improvement in MLP offline skills is likely primarily due to differences in normalization strategies. We used a per-level standard deviation that varies with altitude to make each target variable at each level at the order of 1. The normalization in \citeA{yu2024climsim} used a scaling factor that does not vary vertically, which may prevent effective optimization of targets at all levels. In addition, the choice of the output scaling factors in \citeA{yu2024climsim} made the loss from wind and cloud tendencies too small from the optimizer's perspective compared to the loss from temperature and water vapor tendencies, making their MLP model difficult to learn the wind and cloud tendencies. Additionally, many input features, such as cloud water mixing ratios, have highly skewed distributions with long tails above 10 standard deviations. To address this, we transformed the exponential-like cloud input features into more uniform distributions, as documented in Appendix \ref{appendix:normalization}. Although we did not conduct ablation tests on different normalization strategies, we believe that properly transforming skewed distributions with extreme tails to less skewed distributions (ideally normal distribution) can enhance training dynamics and improve model performance.

Switching from MLP to U-Net-baseline significantly improves the R$^2$ for almost all target variables at all levels, suggesting that the U-Net architecture is more expressive and better suited to capture the data compared to the baseline MLP model. Expanded input features further improve the R$^2$ values in the U-Net-expanded model for almost all target variables at all levels when compared to the U-Net-baseline model. Although we do not present a formal ablation of the individual effects of large-scale forcing and convection memory, preliminary tests indicate that both components contribute to improving the R$^2$ scores.

When comparing the U-Net models with and without the microphysical constraint, the R$^2$ values are quite similar overall. However, the U-Net model with the liquid-ice partitioning constraint shows a marked improvement in predicting liquid cloud tendencies at several levels above about 350 hPa (Figure \ref{r2-offline}c). At these levels, the constrained U-Net shows very high R$^2$ values, while the unconstrained U-Net shows little skill. These levels correspond to temperatures too cold for liquid clouds to exist, where the CRM tends to remove any liquid clouds that may have been advected there by the dynamical core. We also observe a decrease in R$^2$ for water vapor tendencies above the tropopause in the U-Net-expanded-constrained model (Figure \ref{r2-offline}b), though the cause remains unclear. One possible explanation is that using only the total cloud mixing ratio as input may omit relevant information needed to accurately predict water vapor tendencies at these levels, although it is impossible to rule out that other factors are at play.

\begin{figure}
\includegraphics[width=1\textwidth]{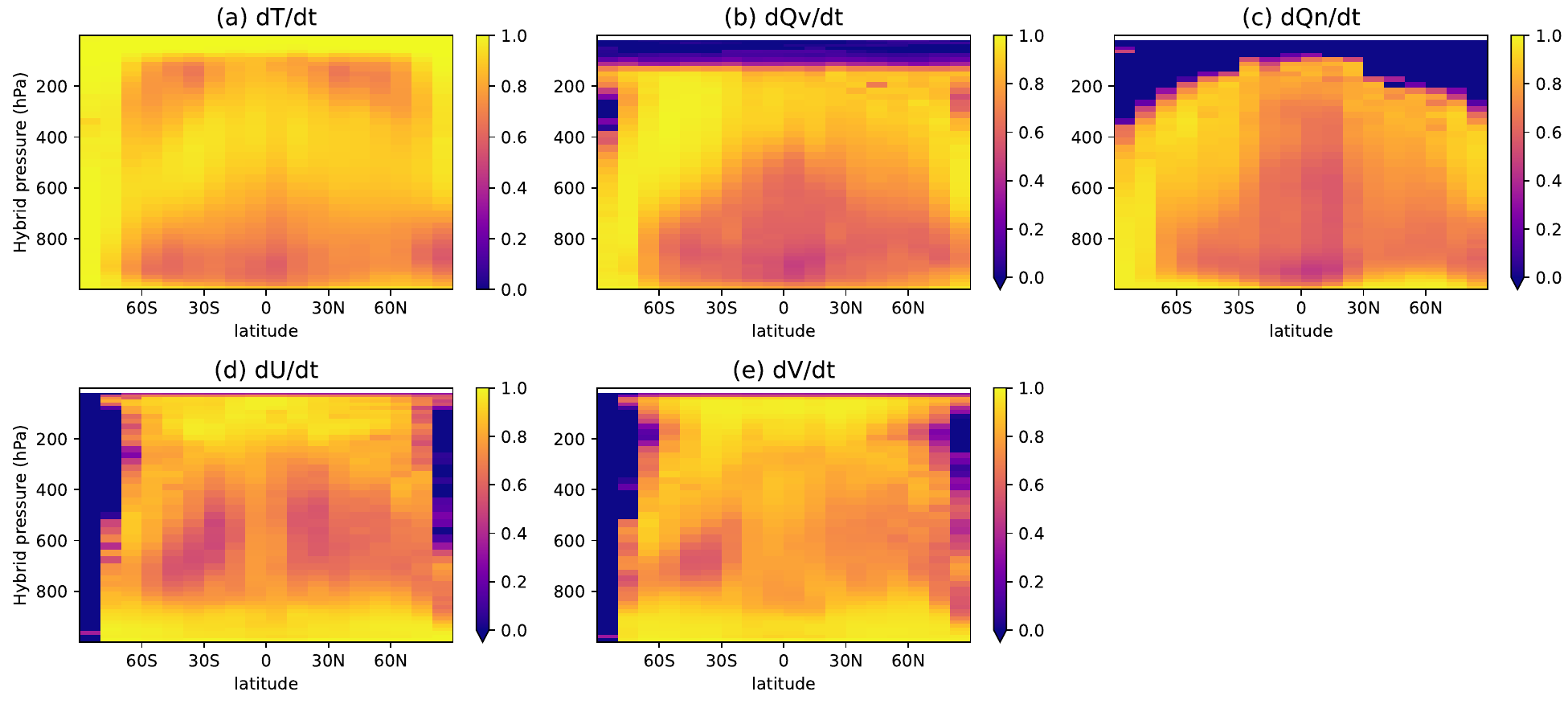}
\centering
\caption{
Zonal mean R$^2$ evaluated on the test set for the U-Net-expanded-constrained model. The five panels denote the R$^2$ for the tendencies of temperature (a), water vapor (b), total cloud (liquid plus ice) mixing ratio (c), zonal wind (d), and meridional wind (e).
}
\label{offline-r2}
\end{figure}

Figure \ref{offline-r2} shows the zonal mean R$^2$ for different variables for the U-Net-expanded-constrained model. Zonal mean R$^2$ is evaluated by first calculating the R$^2$ for each column across samples at different times and then averaging the R$^2$ zonally. The negative R$^2$ of vapor and cloud at the stratosphere is likely because the vapor and cloud tendencies at those high levels are too small and noisy to predict. Our U-Net model also poorly predicts wind tendencies at the north and south poles. More careful normalization or more training data might be required to improve the R$^2$ there.

\section{Online results}
\label{section:online}

\subsection{Online Error Development in the Initial First Year}
\label{section:online-1year}

\begin{figure}
\includegraphics[width=1\textwidth]{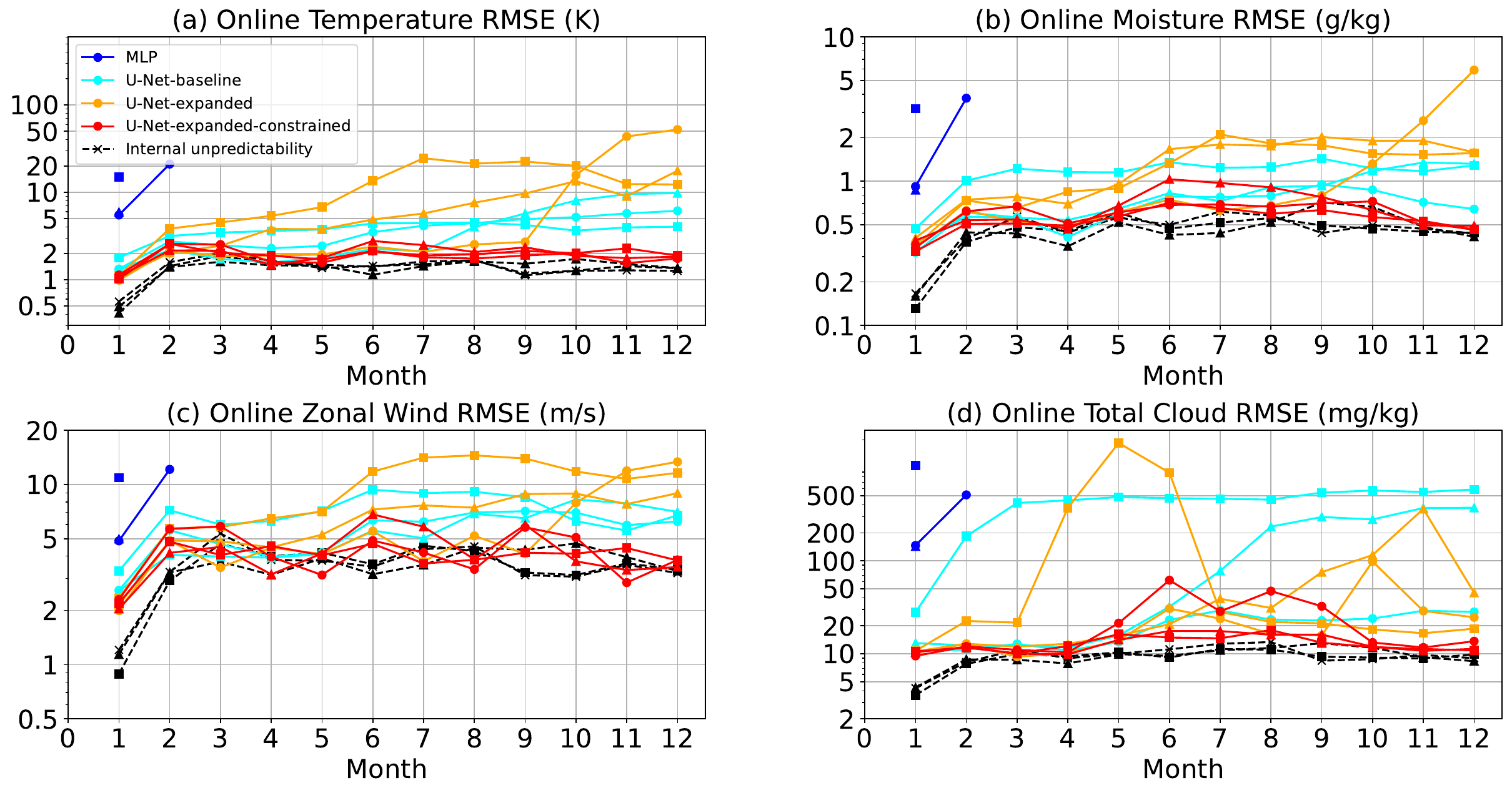}
\centering
\caption{
Online monthly and globally averaged (both horizontally and vertically and weighted by mass in each grid) RMSE of temperature (K), water vapor (g/kg), zonal wind (m/s), and total cloud mixing ratio (mg/kg) over a one-year period, comparing baseline MLP (blue), U-Net-baseline (cyan), U-Net-expanded (orange), and U-Net-expanded-constrained (red) against the reference E3SM-MMF simulation. The dashed black lines represent the uncertainty of the reference physical simulation due to internal atmospheric variability. Each architecture was trained with three slightly different configurations (e.g., loss functions and learning rate schedules) to sample ML uncertainty, represented by different markers but with the same line color.
}
\label{timeseries-1y}
\end{figure}

While the offline skill is important, the more critical aspect of the hybrid simulations is the online performance. No previous work under the MMF framework has achieved good online error with full condensate coupling. Figure \ref{timeseries-1y} summarizes the online performance of the baseline MLP models and various U-Net models. All the hybrid simulations and the reference E3SM-MMF simulations in Section \ref{section:online} use an identical initial condition. 

For each hybrid simulation, we evaluate the monthly evolution of the globally averaged RMSE of monthly-mean temperature, water vapor, zonal wind, and total cloud mixing ratio over the course of one-year hybrid simulations. The RMSE quantifies the discrepancy of monthly-mean fields between the hybrid simulations and the reference E3SM-MMF simulation. The exact definition of the monthly RMSE is in \ref{appendix:online_metric}. Here we focus on monthly-mean values rather than instantaneous values because the instantaneous weather pattern is chaotic and becomes largely unpredictable beyond the initial autocorrelation timescale. In contrast, monthly-mean patterns associated with seasons and geography remain more predictable. 

We ran a small ensemble of E3SM-MMF simulations to estimate the internal atmosphere unpredictability (dashed black lines in Figure \ref{timeseries-1y}), representing an error floor that the hybrid simulations should not be expected to surpass (i.e., the performance of a perfect emulator). The E3SM-MMF climate model runs on a hybrid CPU/GPU system and is not bit-for-bit reproducible. Tiny numerical differences due to the non-associativity of floating point arithmetic and the use of atomic operations (specific to how the current E3SM-MMF version implements parallel reductions) in the calculations can grow over time, leading to small but non-zero uncertainty even in globally averaged metrics of monthly atmospheric states in pure physical simulations. Such numerical differences, when accumulated over a few steps, naturally act as perturbations to the initial conditions that generate ensemble simulations. To create the ensemble, we ran the reference E3SM-MMF climate model three additional times using identical initial conditions and compared the resulting differences to the reference simulation.

The simulations using MLP models all crashed within the first three months and exhibited large RMSE in the first month. In contrast, all the U-Net simulations allowed for more stable integrations, with all U-Net simulations completing the full year. For the first simulated month, where all hybrid runs finished, the U-Net simulations demonstrated significantly lower monthly RMSE of all variables compared to the hybrid simulations using the MLP models.

\begin{figure}
\includegraphics[width=1\textwidth]{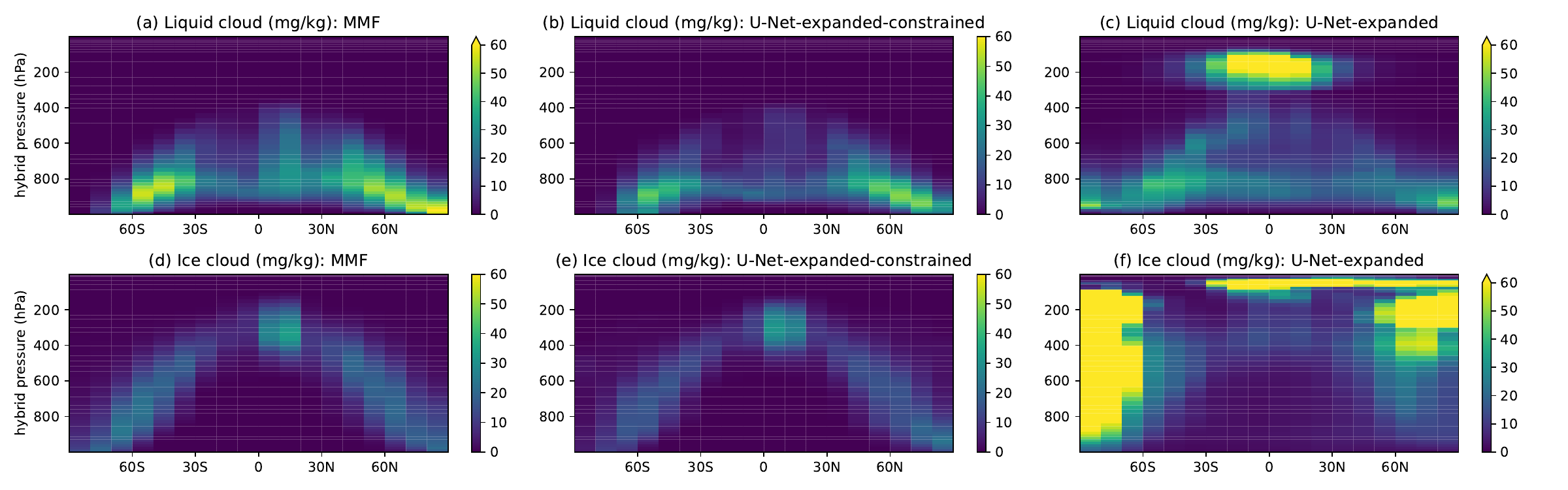}
\centering
\caption{
The zonal mean liquid (upper row) and ice (lower row) cloud mixing ratio at month 10 in the reference E3SM-MMF simulation (left) and in the hybrid simulations with the U-Net-expanded-constrained model (middle) and with the unconstrained U-Net-expanded model (right).
}
\label{ablation-microphysics}
\end{figure}

However, the U-Net models can still experience large online error growth. For the U-Net models without microphysical constraints (U-Net-baseline and U-Net-expanded), while their first-month RMSE is reasonably satisfactory, the online error becomes unacceptably large even a few months into the hybrid simulation (orange and cyan lines in Figure \ref{timeseries-1y}), e.g., with temperature RMSE exceeding 10 K—far surpassing the level of internal unpredictability. Such large RMSE suggests a systematic bias in the hybrid simulation that is significantly larger than the variability of the monthly climatology. Indeed, this trend is reflected in Figure S1 of the Supplementary Information, which shows the monthly evolution of global mean near-surface temperature, precipitable water, and total cloud path in the hybrid and reference simulations. Both the U-Net-baseline and U-Net-expanded simulations significantly diverge from the reference E3SM-MMF simulations after the first year, consistent with the observed RMSE growth.


We hypothesize that a root cause of the error growth is unrealistic and insufficiently constrained cloud formation. The first line of evidence is that the unconstrained U-Net models can have greater than 100 mg/kg RMSE of total clouds in some months (Figure \ref{timeseries-1y}d). Figure \ref{ablation-microphysics} compares the zonal mean liquid and ice cloud mixing ratios between the reference pure physics simulation and the U-Net-expanded-constrained and U-Net-expanded simulations at month 10, i.e. when one of the U-Net-expanded simulations (the orange line with the circle markers in Figure \ref{timeseries-1y}) begins to show highly unrealistic cloud formation and quick growth of the RMSE of temperature and water vapor. 

The second, most telling, line of evidence is that the unphysical cloud phase is immediately apparent as a pathology in the hybrid condensate climatology. The unconstrained U-Net hybrid run shows an excessive amount of liquid cloud near the tropical tropopause (Figure \ref{ablation-microphysics}c), where the temperature is too cold to allow any liquid water. We also observe excessive cloud ice in the stratosphere (Figure \ref{ablation-microphysics}f). These unrealistic cloud distributions are clearly out-of-sample conditions that never existed in the training data, and the unconstrained U-Net fails to remove them, continuing to make extrapolated predictions based on these out-of-sample input features. This out-of-distribution challenge only becomes apparent in the online evaluation of the hybrid models and is difficult to reveal with just offline evaluation.

Conversely, when we directly enforce the microphysical constraints on the liquid-ice partition and constrain the existence of deep penetrating overshoots, the zonal mean cloud distribution appears more consistent with the pure physics simulation (Figure \ref{ablation-microphysics}b and e vs. a and d). With these constraints, the downstream RMSE stabilized after an initial rise within the first two months. The global temperature RMSE remained around 2K, and the global water vapor RMSE stayed below 1 g/kg (Figure \ref{timeseries-1y}a and b). The global zonal wind RMSE remains around 5 m/s, and the global total cloud RMSE remains within 20 mg/kg except for one of the checkpoints showing an RMSE of 50 mg/kg in summer months.

As a substantial advancement, the U-Net-expanded-constrained models begin to match the monthly global mean climatology as well as the global variability in the reference E3SM-MMF simulations (Figure S1 and S2 of the Supplementary Information). Although the online RMSE of the hybrid simulations is generally slightly higher than the internal unpredictability (roughly within 1K temperature RMSE and 0.5 g/kg water vapor RMSE), in some months, the RMSE of the hybrid simulations overlaps with the internal unpredictability levels (Figure \ref{timeseries-1y}), suggesting the systematic bias of the U-Net-expanded-constrained simulations becomes comparable to the variability of the monthly climatology.

\subsection{Climatology Evaluation of Stable 5-year Hybrid Simulations}
\label{section:online-5year}

\begin{figure}
\includegraphics[width=1\textwidth]{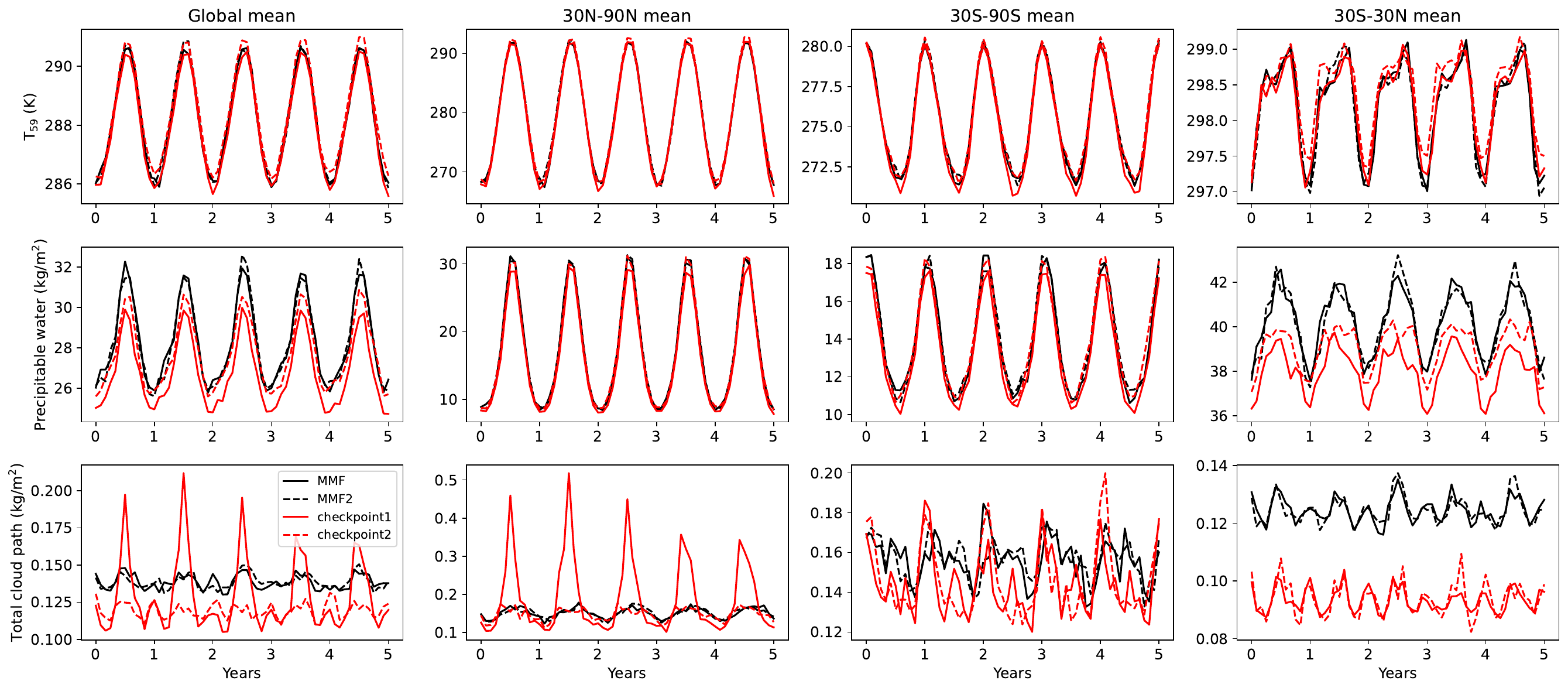}
\centering
\caption{
Five-year time series of area-weighted mean values for near-surface temperature (at the lowest model level, top row), precipitable water (middle row), and total cloud path (sum of liquid and ice, bottom row). The data is shown across four regions: global mean (first column), 30°N–90°N (second column), 30°S–90°S (third column), and 30°S–30°N (fourth column). The solid black line and dashed black line represent two reference 5-year E3SM-MMF simulations initialized with the same initial conditions to sample internal variability (similar to the dashed lines in Figure \ref{timeseries-1y}). The solid and dashed red lines represent hybrid simulations using two U-Net-expanded-constrained checkpoints: Huber loss with reduce-on-plateau scheduler (solid red) and Huber loss with step scheduler (dashed red) (see \ref{appendix:training}).
}
\label{timeseries-5y}
\end{figure}

The U-Net-expanded-contrained simulations can be stably integrated for at least 5 years (Figure \ref{timeseries-5y}). Due to the maximum job runtime of 24 hours on the computing cluster we used, we only ran the hybrid simulations for five years plus one month and evaluated the 5-year mean climatology after dropping the first month for the spin-up period. Restart capabilities are not yet engineered and are nontrivial to implement given that the ML model collects a non-standard history from multiple previous time steps. However, we expect our hybrid simulations can integrate longer when there is no maximum job runtime limit. 

During the 5-year simulations, the global mean near-surface air temperature, precipitable water, and total cloud (liquid plus ice) path reasonably reproduce the seasonal cycle of the reference E3SM-MMF simulation without any significant drift (Figure \ref{timeseries-5y}). Two hybrid simulations using different U-Net-expanded-constrained checkpoints are shown to distinguish between systematic and checkpoint-dependent errors. The reference physics simulation was run twice for five years using identical initial conditions (black solid and dashed lines) to sample internal unpredictability and represent what a perfect emulation can achieve. Figure S2 in the Supplementary Information further demonstrates that the U-Net-expanded-constrained simulation generally captures the seasonal variation of spatial variability of near-surface air temperature and precipitable water seen in the reference E3SM-MMF simulation. However, during the northern hemisphere summer, the hybrid simulation exhibits slightly higher variance in precipitable water for the region spanning 40°N–50°N compared to the reference E3SM-MMF simulation. 

Systematic biases still remain. The hybrid U-Net-expanded-constrained simulations displayed systematic underestimations of precipitable water and total cloud path, primarily in the tropics (Figure \ref{timeseries-5y}). One checkpoint exhibited a substantial overestimation of the total cloud path in the northern hemisphere (solid red line), while the other did not (dashed red line).

\begin{figure}
\includegraphics[width=0.8\textwidth]{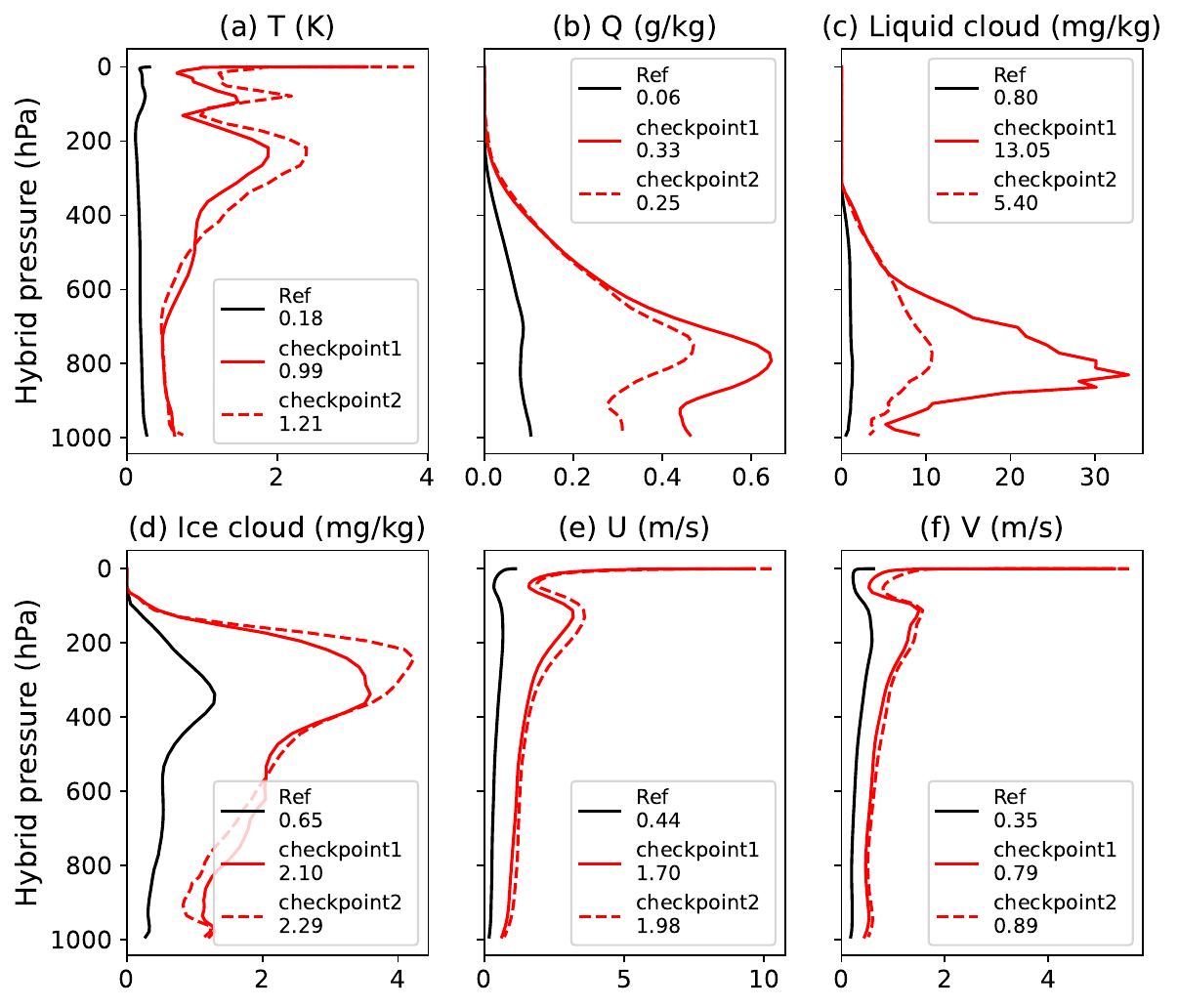}
\centering
\caption{
Online RMSE of the 5-year time-mean climatology for temperature (a), water vapor (b), liquid cloud mixing ratio (c), ice cloud mixing ratio (d), zonal wind (e), and meridional wind (f). Profiles represent RMSE values per vertical level, while the global RMSE values (integrated across horizontal and vertical space) are noted in the legend. RMSE is defined as in \ref{appendix:online_metric}. The black line shows the RMSE between two reference E3SM-MMF simulations with identical initial conditions, serving as a proxy for a perfect emulator (diverging only due to non-bit-for-bit numerical differences). The solid and dashed red lines represent the online RMSE of hybrid simulations using two U-Net-expanded-constrained checkpoints: Huber loss with a reduce-on-plateau scheduler (solid red) and Huber loss with a step scheduler (dashed red) (see \ref{appendix:training}).
}
\label{5year-rmse-profiles}
\end{figure}

Figure \ref{5year-rmse-profiles} summarizes the RMSE of the 5-year time-mean climatology for state variables across vertical levels, along with the 3D global mean values (noted in the legend). Both U-Net-expanded-constrained checkpoints achieved a time-mean temperature RMSE below 2K and water vapor RMSE below 1 g/kg throughout the troposphere. The global mean RMSE for temperature and water vapor was 0.99K and 0.33 g/kg for one checkpoint, and 1.21K and 0.25 g/kg for the other. The checkpoint with better cloud condensate performance (dashed red line) exhibited a peak liquid cloud RMSE of 10 mg/kg near 800 hPa and an ice cloud RMSE of 4 mg/kg near 200 hPa. RMSE values for the zonal and meridional winds remained within 2 m/s and 1 m/s, respectively, across most levels, with exceptions at the uppermost level.

\begin{figure}
\includegraphics[width=1\textwidth]{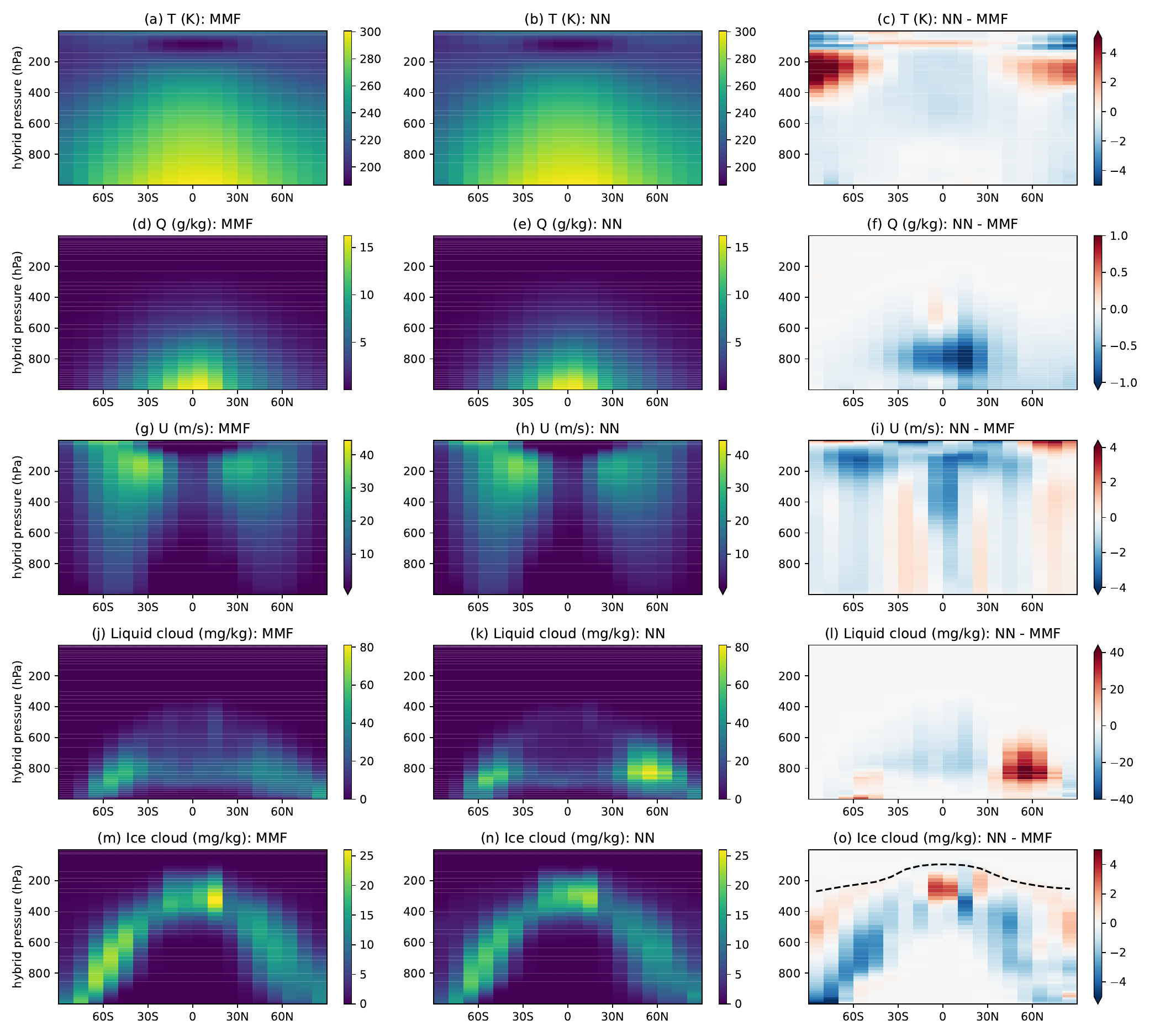}
\centering
\caption{
Five-year zonal mean atmospheric state in the reference E3SM-MMF simulation (left) and in the hybrid simulation with the U-Net-expanded-constrained model (middle). The right column shows the zonal mean bias as the mean state from the hybrid simulation minus that of the reference simulation. The five rows show temperature, water vapor, zonal wind, liquid cloud, and ice cloud. The Black dashed line in panel o indicates the 5-year mean position of the tropopause level used in the microphysical constraints.
}
\label{online-states}
\end{figure}

Figure \ref{online-states} shows the zonal mean bias of the 5-year hybrid simulation climatology, with a tropospheric zonal mean temperature bias below 2K and a water vapor bias below 1 g/kg. Figure S3 in the Supplementary Information shows that the zonal mean relative humidity bias is generally within 4\% across most of the troposphere, with a peak bias of 8\% in the tropical regions. However, larger relative humidity biases are observed in extremely cold environments, such as near the surface at the South Pole and in the stratosphere. Additionally, we control the global zonal mean zonal wind bias within 5 m/s and maintain a reasonable distribution of liquid and ice clouds. While we have a near 40 g/kg cloud water bias near 60N with this model checkpoint, another different checkpoint has only within 10 g/kg cloud water bias in most places (Figure \ref{online-states-another-checkpoint} in \ref{appendix:online-bias-another-checkpoint}).

Our performance surpasses the online zonal mean temperature and water vapor biases reported by \citeA{han2023ensemble}. \citeA{han2023ensemble} showed greater than 10K temperature bias and near 1.5 g/kg water vapor bias in the polar regions in their hybrid simulations (see Figure S9 in their Supplementary Information). \citeA{han2023ensemble} did not provide the online bias of their wind and cloud climatology so we cannot make a comparison on these variables. 


\begin{figure}
\includegraphics[width=1\textwidth]{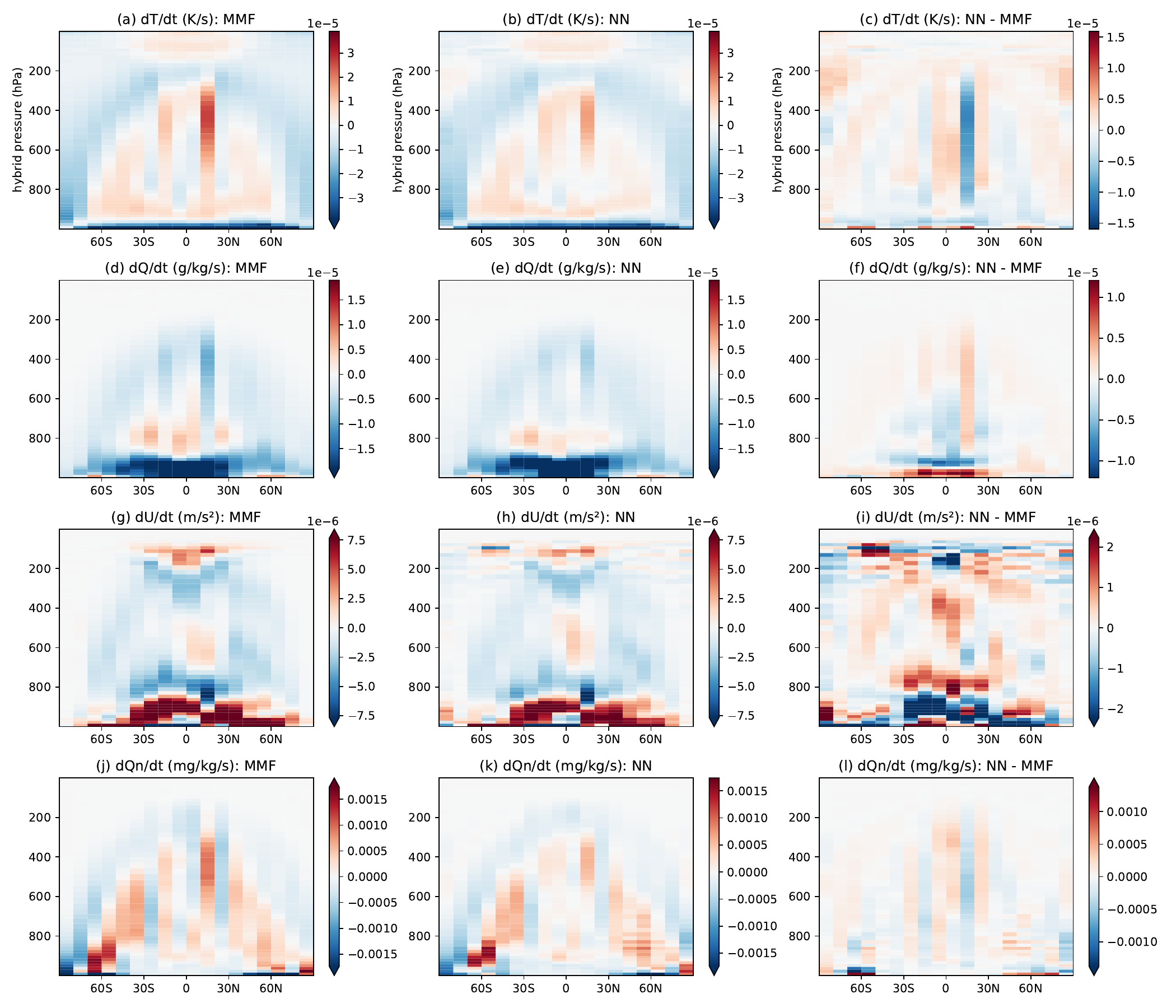}
\centering
\caption{
Five-year zonal mean atmospheric tendencies in the reference E3SM-MMF simulation (left column), in the hybrid online simulation with the U-Net-expanded-constrained model (middle column), and the difference (hybrid minus MMF run, right column). The four columns correspond to the tendencies of temperature, water vapor, zonal wind, and total cloud (liquid plus ice) mixing ratio. 
}
\label{online-tendency}
\end{figure}

Figure \ref{online-tendency} presents the five-year averaged, zonal mean subgrid tendencies of temperature, water vapor, zonal wind, and total cloud mixing ratio for both the reference E3SM-MMF simulation and the hybrid run with the constrained U-Net model. The hybrid simulation effectively reproduces the patterns of these tendencies. The temperature and water vapor tendencies in the reference simulation are well captured by the hybrid simulation, showing similar vertical and meridional distributions, such as the two heating and drying bands in the tropics corresponding to regions with frequent deep convection and precipitation. Additionally, the hybrid simulation successfully replicates the zonal wind and total cloud mixing ratio tendencies, demonstrating its ability to correctly capture the large-scale dynamics in the E3SM-MMF physical simulation. However, some discrepancies exist between the hybrid simulation and the reference E3SM-MMF simulation. For instance, the hybrid simulation underestimates the heating and drying tendency band near 15N, underestimates the moistening tendency in the tropical lower troposphere, and produces anomalous heating in the polar stratosphere near 200 hPa.

\begin{figure}
\includegraphics[width=1\textwidth]{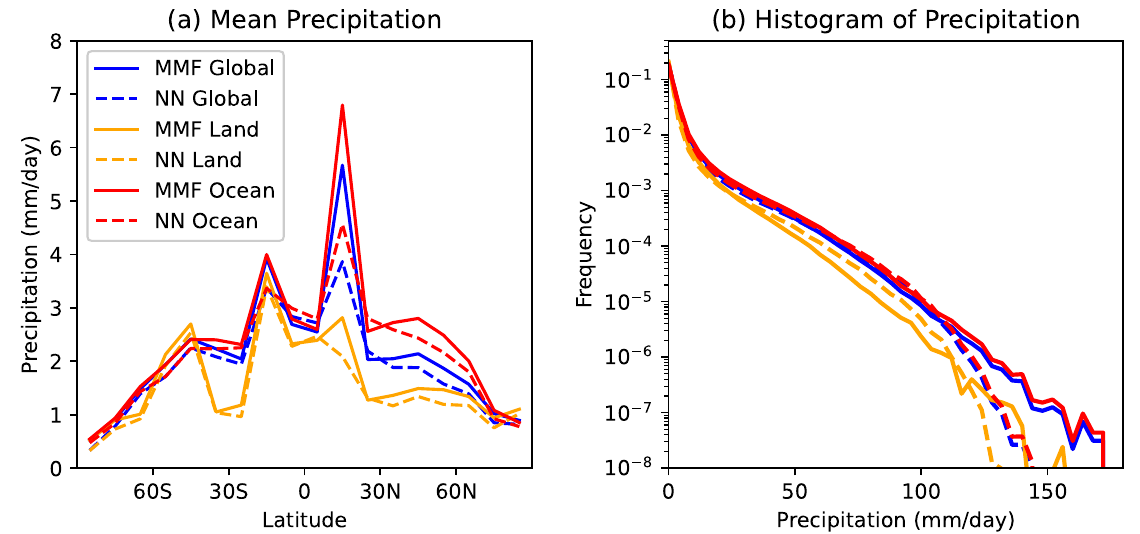}
\centering
\caption{
(a) 5-year zonal mean precipitation in the reference E3SM-MMF simulation (solid lines) versus the online hybrid simulation with the U-Net model incorporating microphysical constraints (dashed lines). Different colors denote samples taken globally (blue), only over land (orange), and only over the ocean (red). (b) Histogram of the hourly instantaneous precipitation.
}
\label{precip}
\end{figure}

\begin{figure}
\includegraphics[width=1\textwidth]{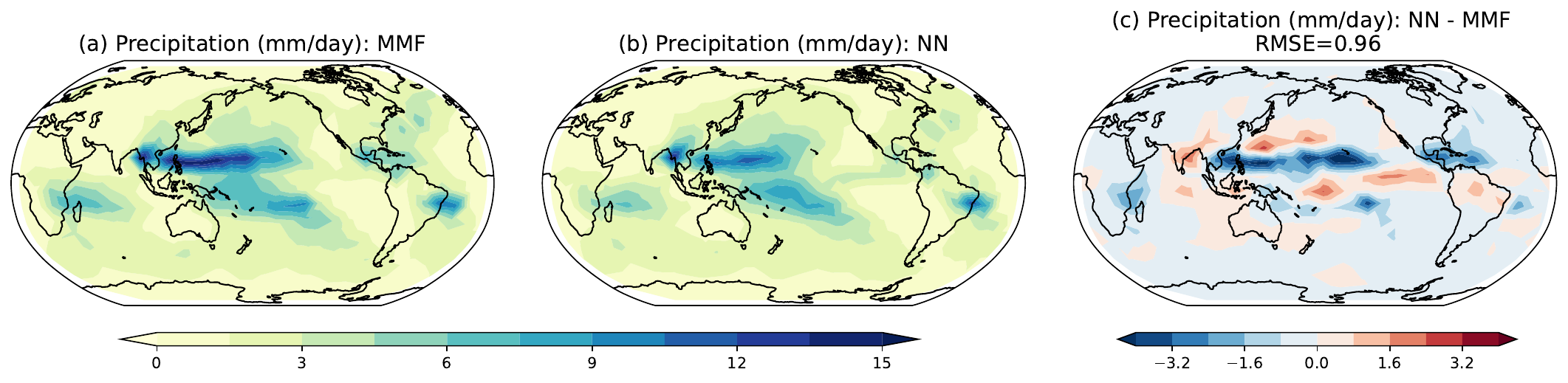}
\centering
\caption{
Horizontal maps of the five-year mean surface precipitation: (a) reference E3SM-MMF simulation, (b) hybrid U-Net-expanded-constrained simulation, and (c) bias (hybrid minus reference). Panel c also documents the RMSE of 0.96 mm/day for the five-year mean precipitation between the hybrid and reference simulations.
}
\label{precip-map}
\end{figure}

The online precipitation statistics in the 5-year hybrid simulation reasonably reproduced the reference E3SM-MMF simulation (Figure \ref{precip}). The zonal mean, 5-year average precipitation distribution in the hybrid simulation matches that of the reference E3SM-MMF simulation (Figure \ref{precip}a). However, the precipitation peak in the northern hemisphere is underestimated, primarily over the ocean. Figure \ref{precip-map} presents the horizontal map of the 5-year mean precipitation pattern. The hybrid simulation captures the spatial distribution of global precipitation reasonably well but underestimates the rain belt on the northern side of the equator. Quantitatively, the hybrid simulation exhibits an RMSE of 0.96 mm/day for the 5-year mean precipitation.

The hybrid run also reasonably reproduces the extreme tail of the PDF of the instantaneous precipitation sampled every hour (Figure \ref{precip}b). The frequency in the hybrid run aligns well with the reference simulation down to a frequency of 10$^{-6}$, demonstrating similar skills that match previous studies \cite{rasp2018deep,yuval2020stable,wang2022stable}. However, beyond the frequency of 10$^{-6}$, the hybrid simulations failed to reproduce the most extreme and rarest of precipitation events.

\section{Conclusions}

The Multi-scale Modeling Framework (MMF) offers a promising solution to better represent subgrid processes that are poorly parameterized in traditional climate models. By integrating machine learning (ML) to emulate the embedded cloud-resolving models, we can potentially reduce computational costs substantially while maintaining high fidelity in the resulting hybrid climate simulations. Compared to other hybrid approaches, the MMF framework provides the convenience of locality due to its scale separation and retains all necessary components (e.g., ocean and land modules, aerosol module) for future climate change projections. 

Here, we used the coarse-resolution version (with an effective resolution of 11.5°$\times$11.5°) of the ClimSim data \cite{yu2024climsim} to train a stable ML parameterization that emulates the full physics (convection plus radiation) of the cloud-resolving model in the E3SM-MMF climate model. In the hybrid MMF simulation studies, to our knowledge, this is the first work to achieve a multi-year stable ML parameterization that emulates the full set of variables required to couple with the dynamical core and surface models with geography. Our hybrid simulation achieved state-of-the-art online 5-year zonal-mean biases in temperature and water vapor climatology compared to previous hybrid MMF studies with similar complexities \cite{han2023ensemble,behrens2024improving}. Our hybrid simulation also achieved realistic multi-year cloud climatology with explicit cloud condensate coupling. 

We utilized an improved U-Net architecture with expanded input features, such as large-scale forcings and convective memory, which significantly improved offline fitting skills compared to the baseline multi-layer perceptron architecture used in many previous studies. While offline skill is not completely representative of online performance, we found that incorporating microphysical constraints in the ML prediction significantly enhanced the online performance of the hybrid simulations. These constraints were informed by the initial analysis of hybrid errors in the simulated condensate. We found that they help maintain a reasonable liquid and ice cloud distribution in the hybrid simulation. They also significantly benefited other state variables such as temperature and water vapor bias through associated dynamical couplings. Our microphysics-constrained U-Net model also reasonably reproduced the spatial distribution of five-year mean precipitation and the frequency of extreme precipitation compared to the reference E3SM-MMF simulation, with a root mean square error of 0.96 mm/day for five-year mean precipitation.

Despite these advancements, our multi-year online performance reveals systematic biases that remain to be addressed. For example, we observe a persistent low-bias in tropical lower troposphere water vapor and liquid cloud mixing ratio, as well as a vertical dipole pattern in temperature biases within the polar stratosphere. Additionally, the hybrid simulation underestimates precipitation rates over the northern flank of the equatorial rain belt. Addressing these biases presents significant challenges, underscoring the need for innovative approaches to optimizing online performance and improving the fidelity of hybrid climate simulations.

Our work demonstrates the potential of achieving stable and skillful ML parameterization for operational-level hybrid climate simulations. We illustrate the importance and potential ways of incorporating domain knowledge and physical constraints into the ML parameterization to improve the online bias and stability of hybrid physics-ML climate simulations. We prove that including explicit cloud condensate coupling in our ML parameterization is achievable while maintaining satisfactory multi-year climatology, which would be crucial to further increase the complexity of the ML parameterization such as including pollutant transport and cloud-aerosol interaction.

\section{Discussion and Limitations}

\subsection{Complexity of the Reference and Hybrid Simulations}
Our online results establish a robust baseline that approaches operational-level complexity, demonstrating substantial progress, but also highlighting areas needing further development. Looking forward, additional complexity is required in the ML parameterization and the training data to achieve complete compatibility for the purpose of climate projection. For example, our online simulations used prescribed sea surface temperature and sea ice. To simulate future climate change, it will be necessary to train and evaluate the hybrid simulations with a fully coupled ocean module that allows for interactions between the atmosphere, ocean, and sea ice. Aerosol-cloud interaction is also a critical process for climate projection but is not included in the simulations that generate the ClimSim data. To develop ML parameterization that includes aerosol-cloud interaction, one can create a new training dataset that includes aerosol-related inputs and outputs. This can be done using simulations that implement the Explicit-Cloud Parameterized-Pollutant (ECPP) scheme \cite{gustafson2008explicit}.

For computational expedience, our ML parameterization is trained on data from the coarse-resolution E3SM-MMF simulations in the ClimSim dataset \cite{yu2024climsim}, with an effective resolution of 11.5°$\times$11.5°, and tested in a similarly coarse prognostic setting. On the one hand, this serves as a simple, cost-efficient starting point to rapidly iterate on different ideas to improve offline and online performance. However, this resolution poorly simulates tropical precipitation, synoptic scale flows, fronts, and large-scale convective systems like Madden-Julian Oscillation (MJO). The coarse resolution introduces challenges for the parameterization problem, as it requires the parameterization to account for not only unresolved convection and clouds but also a substantial portion of the large-scale circulation typically resolved in an atmospheric model. However, it may also simplify the problem by avoiding the “gray zone” issues encountered in models with resolutions of 50–100 km, where some gravity waves, fronts, and large-scale convective systems are partially resolved but poorly simulated.

Future work should thus extend this study to the most operationally relevant, highest-resolution data in the ClimSim dataset (effective resolution of 1.5°$\times$1.5°), which better resolves synoptic-scale flows and contains more detailed topography. However, higher-resolution data may shift input distributions significantly, including changes in mean atmospheric states, large-scale advective forcings, and the emergence of previously unencountered extrema, such as newfound peak topography and new land surface properties. These differences will likely require retraining the ML parameterizations to handle the altered data characteristics and ensure reliable performance in the high-resolution setting.

\subsection{Improving the Offline and Online Performance}

Our results are consistent with the view of \citeA{lin2023sampling} that pursuing higher offline skills can lead to improved online performance in hybrid simulations, suggesting exploring advanced architectures might be beneficial. MLPs with deeper architectures (10+ layers) and residual connections \cite{kochkov2024neural,wang2022stable} may offer greater expressive power compared to the MLP model evaluated in this study and may exhibit different online performance. Transformer models or Long Short-Term Memory (LSTM) models might effectively learn vertical structure of the input features. A Recurrent Neural Network (RNN)-style model in the time dimension, like a state-space model or LSTM, could better represent the convection memory.

In addition to exploring advanced architectures, more careful normalization in the upper troposphere and the stratosphere or location-dependent normalization might further improve offline skills. While these advancements hold promise in general, our current U-Net model and normalization strategies continue to face significant challenges in specific areas. It struggles to predict stratospheric cloud and water vapor tendencies as well as wind tendencies in the polar regions. 

For further insights into improving offline performance, readers are encouraged to review the solutions from a recent Kaggle competition, “LEAP-Atmospheric Physics AI (ClimSim)” (https://www.kaggle.com/competitions/leap-atmospheric-physics-ai-climsim). The participants worked on the same ClimSim dataset and explored various innovative architectures, data preprocessing techniques, and training strategies to optimize the offline skill of their models. For example, we speculate that the low offline R$^2$ of water vapor, ice cloud, and wind tendencies in the tropopause and stratosphere (Figure \ref{r2-offline}) could be improved by better normalization and architectures.

It could be viewed as surprising that the expanded input features (large-scale advective forcing plus previous steps' convection memory) did not improve the online performance (Figure \ref{timeseries-1y}) given that some previous studies, such as \citeA{lin2023sampling}, have suggested adding convection memory can improve stability and online performance. A potential explanation is that a larger sample size of architectures may be necessary to robustly assess the impact of these expanded input features. We encourage future studies to further test and ablate these features to better understand their effects on model performance.

Exploring the pathologies of bias and instability in hybrid climate simulations is highly beneficial. In our study, the observation of unrealistic cloud formation was crucial in motivating our implementation of microphysical constraints. These unrealistic clouds (clearly out-of-sample simulation states) highlight a broader challenge in hybrid physics-ML simulations: ensuring that ML-generated physics do not drive the simulation into non-physical states. Addressing this issue is difficult because training data sourced solely from physically accurate numerical simulations may not prepare the ML model to correct such errors. Our work demonstrates that addressing such out-of-sample issues can benefit from in-depth error diagnosis and constraints informed by domain knowledge to prevent un-physical situations from happening in the first place.

Our hybrid simulations exhibit certain bias patterns, such as the vertical dipole temperature bias in the polar stratosphere (Figure \ref{online-states}c), a drying water vapor bias, and a liquid cloud low bias in the tropical lower troposphere (Figure \ref{online-states}f and \ref{online-states}l). These bias patterns also exist in the hybrid simulations across all the checkpoints (e.g., Figure \ref{online-states-another-checkpoint}) of the U-Net model with microphysical constraints. \ref{appendix:error-growth} shows that the tropical lower troposphere bias patterns of too little water vapor and liquid cloud develop very quickly within the first few days and maintain their magnitude for years. 

Some other bias patterns vary with different checkpoints. The constrained U-Net model checkpoint used to create Figure \ref{online-states} corresponds to the red circle line in Figure \ref{timeseries-1y}, which exhibits slightly better temperature and water vapor zonal mean biases but worse biases for online cloud mixing ratio than other U-Net checkpoints (Figure \ref{timeseries-1y}d). The positive bias of the lower troposphere liquid cloud (Figure \ref{online-states}l) can be significantly improved with other checkpoints (Figure \ref{online-states-another-checkpoint}). A more extensive hyperparameter search could potentially further optimize some of the observed online bias.

Further optimizing the online bias of our constrained U-Net model can be challenging once an optimal architecture is decided upon. Unlike offline errors, online errors are non-differentiable and cannot be directly optimized using gradient descent. Once the model architecture is fixed, we currently rely on checkpoint searches (e.g., by perturbing training hyperparameters) to identify the checkpoint that yields the best online performance. 

When evaluating online error growth, we used a fixed initial condition for all hybrid and reference simulations. In future work, it would be valuable to better isolate the development of online biases from the sensitivity of reference simulations to initial conditions. For instance, conducting an ensemble of simulations with perturbed initial conditions could provide a clearer measure of the system’s chaotic behavior. Additionally, using different initial conditions across various times of the year could improve the robustness of the evaluation. Analyzing the bias development on daily or even shorter timescales could also offer insights into the unsteady onset of these biases.

It remains an open question how to use gradient-free methods to optimize these online errors. Gradient-free methods such as imitation learning and Ensemble Kalman Inversion could be promising, as they have shown success in other applications \cite{ross2011reduction,rasp2020coupled,kelp2022online,lopez2022training,pahlavan2024explainable,christopoulos2024online}. Another possible yet highly non-trivial approach is to develop a fully differentiable climate model, as demonstrated for the atmospheric primitive equation solver subcomponent by \cite{kochkov2024neural}, although this is a highly nontrivial effort for a fully featured climate model compatible for future climate projection. With a model that can be integrated over time in a differentiable way, it would be possible to optimize multi-step losses using gradient descent. Some of the online error patterns develop very quickly within the first day of the hybrid simulation, and optimizing multi-step losses could potentially help address these error patterns.

There is room for adjustment and improvement in the microphysical constraints we used. The temperature-based liquid-ice partition constraint is not directly applicable to reconfigurations of the CRM that use more advanced microphysics schemes of the sort required for fully explicit aerosol-cloud interaction science. However, some soft constraints can still be applied, such as prohibiting liquid clouds below a certain temperature and similarly prohibiting ice clouds above a warm temperature threshold. The constraint of removing all clouds in the stratosphere eliminates deep penetrating clouds, which are infrequent but not entirely absent. A more careful, data-driven approach could be used to define the tropopause, requiring fewer approximations and better capturing the rare but non-zero deep convective overshoots. Looking ahead, now that this baseline problem that intentionally used simplistic microphysics is proving tractable, extensions to more modern microphysical representations would be timely. 

\section{Open Research}

The training data is the ClimSim dataset \cite{yu2024climsim}. Quickstarts of how to use the ClimSim dataset, data downloading from HuggingFace, data preprocessing, and training codes are available at \url{https://github.com/leap-stc/ClimSim} \cite{climsimrepo}. The numerical atmospheric model that we use to perform pure physical simulations and hybrid simulations is the Energy Exascale Earth System Model (E3SM)-multiscale modeling framework (MMF) \cite{hannah2020initial}. The version of E3SM-MMF that supports pytorch-fortran coupling is openly available at \url{https://github.com/NVlabs/E3SM} \cite{e3sm-model}. We use an open-source library called Pytorch-Fortran \cite{torchfort} to couple our PyTorch models with the E3SM. The Pytorch-Fortran library requires saving the PyTorch model in TorchScript \cite{torchscript}. We use the PyTorch \cite{paszke2019pytorch} and open-source NVIDIA Modulus library \cite{modulus} for training our neural net models. The simulation output data, saved models, and visualization notebooks are available at \url{https://zenodo.org/records/12797810} \cite{zenodo_hu}. 







\appendix

\section{Neural Network Model Configuration}
\label{appendix:nn_configuration}

\subsection{Input and Output Features}
\label{appendix:inpu_output}
\begin{table}[ht]
 \caption{List of input features used in the MLP model.}
 \centering
 \begin{tabular}{l l l l}
 \hline
  \textbf{Variable} & \textbf{Units} & \textbf{Description} & \textbf{Normalization} \\
 \hline
    $T(z)$ & K & Temperature & (x-mean)/(max-min)  \\
    $RH(z)$ &  & Relative humidity   \\
    $q_l(z)$ & kg/kg & Liquid cloud mixing ratio & 1 - exp(-$\lambda x$)  \\
    $q_i(z)$ & kg/kg & Ice cloud mixing ratio & 1 - exp(-$\lambda x$)  \\
    $u(z)$ & m/s & Zonal wind & (x-mean)/(max-min)  \\
    $v(z)$ & m/s & Meridional wind & (x-mean)/(max-min)  \\
    O3$(z)$ & mol/mol & Ozone volume mixing ratio & (x-mean)/(max-min)  \\
    CH4$(z)$ & mol/mol & Methane volume mixing ratio & (x-mean)/(max-min)  \\
    N2O$(z)$ & mol/mol & Nitrous volume mixing ratio & (x-mean)/(max-min)  \\
    PS & Pa & Surface pressure & (x-mean)/(max-min)  \\
    SOLIN & W/m\textsuperscript{2} & Solar insolation & x/(max-min) \\
    LHFLX & W/m\textsuperscript{2} & Surface latent heat flux & x/(max-min)  \\
    SHFLX & W/m\textsuperscript{2} & Surface sensible heat flux & x/(max-min)  \\
    TAUX & W/m\textsuperscript{2} & Zonal surface stress &  (x-mean)/(max-min) \\
    TAUY & W/m\textsuperscript{2} & Meridional surface stress & (x-mean)/(max-min)  \\
    COSZRS & & Cosine of solar zenith angle & (x-mean)/(max-min)  \\
    ALDIF & & Albedo for diffuse longwave radiation & (x-mean)/(max-min)  \\
    ALDIR & & Albedo for direct longwave radiation & (x-mean)/(max-min)  \\
    ASDIF & & Albedo for diffuse shortwave radiation & (x-mean)/(max-min)  \\
    ASDIR & & Albedo for direct shortwave radiation & (x-mean)/(max-min)  \\
    LWUP & W/m\textsuperscript{2} & Surface upward longwave flux & (x-mean)/(max-min) \\
    ICEFRAC & & Sea-ice area fraction   \\
    LANDFRAC & & Land area fraction   \\
    OCNFRAC & & Ocean area fraction   \\
    SNOWHLAND & m & Snow depth over land & (x-mean)/(max-min) \\
 \hline
 \end{tabular}
 \label{input_table}
 \end{table}

 \begin{table}[ht]
 \caption{List of output features used in the MLP and U-Net-baseline model.}
 \centering
 \begin{tabular}{l l l l}
 \hline
  \textbf{Variable} & \textbf{Units} & \textbf{Description} & \textbf{Normalization} \\
 \hline
    $dT(z,t_{0})$ & K/s & Temperature tendency & x/std  \\
    $dq_v(z,t_{0})$ & kg/kg/s & Water vapor tendency & x/std  \\
    $dq_l(z,t_{0})$ & kg/kg/s & Liquid cloud tendency & x/std  \\
    $dq_i(z,t_{0})$ & kg/kg/s & Ice cloud tendency & x/std  \\
    $du(z,t_{0})$ & m/s\textsuperscript{2} & Zonal wind tendency  & x/std \\
    $dv(z,t_{0})$ & m/s\textsuperscript{2} & Meridional wind tendency  & x/std \\
    NETSW & W/m\textsuperscript{2} & Net shortwave flux at surface & x/std  \\
    FLWDS & W/m\textsuperscript{2} & Downward longwave flux at surface & x/std  \\
    PRECSC & m/s & Snow rate (liquid water equivalent) & x/std  \\
    PRECC & m/s & Rain rate & x/std  \\
    SOLS & W/m\textsuperscript{2} & Downward visible direct solar flux to surface  & x/std \\
    SOLL & W/m\textsuperscript{2} & Downward near-IR direct solar flux to surface & x/std  \\
    SOLSD & W/m\textsuperscript{2} & Downward visible diffuse solar flux to surface & x/std  \\
    SOLLD & W/m\textsuperscript{2} & Downward near-IR diffuse solar flux to surface & x/std  \\
 \hline
 \end{tabular}
 \label{output_table}
 \end{table}

 \begin{table}[ht]
 \caption{List of additional input features used in the U-Net-expanded model and the U-Net-expanded-constrained model. The last three rows are only used for the U-Net-expanded-constrained model. In addition, the U-Net-expanded-constrained model excludes $q_l(z)$, $q_i(z)$, $dq_l(z,t_{-1},t_{-2})$, and $dq_i(z,t_{-1},t_{-2})$ from the inputs.}
 \centering
 \begin{tabular}{l l l l}
 \hline
  \textbf{Variable} & \textbf{Units} & \textbf{Description} & \textbf{Normalization} \\
 \hline
    $dT_{adv}(z,t_0,t_{-1})$ & K/s & Large-scale forcing of temperature & x/(max-min)  \\
    $dq_{T,adv}(z,t_0,t_{-1})$ & kg/kg/s & Large-scale forcing of total water & x/(max-min) \\
    $du_{adv}(z,t_0,t_{-1})$ & m/s\textsuperscript{2} & Large-scale forcing of zonal wind & x/(max-min)  \\
    $dT(z,t_{-1},t_{-2})$ & K/s & Temperature tendency  & x/std \\
    $dq_v(z,t_{-1},t_{-2})$ & kg/kg/s & Water vapor tendency & x/std  \\
    $dq_l(z,t_{-1},t_{-2})$ & kg/kg/s & Liquid cloud tendency &  x/std  \\
    $dq_i(z,t_{-1},t_{-2})$ & kg/kg/s & Ice cloud tendency &  x/std  \\
    $du(z,t_{-1},t_{-2})$ & m/s\textsuperscript{2} & Zonal wind tendency & x/std  \\
    cos(lat) & & Cosine of latitude   \\
    sin(lat) & & Sine of latitude   \\
    $liq\_partition(z)$ &  & Fraction of liquid cloud   \\
    $q_n(z)$ & kg/kg & Total cloud (liquid + ice) mixing ratio & 1 - exp(-$\lambda x$)  \\
    $dq_n(z,t_{-1},t_{-2})$ & kg/kg/s & Total cloud tendency &  x/std  \\
 \hline
 \end{tabular}
 \label{additional_input_table}
 \end{table}
 
Table \ref{input_table} and \ref{output_table} list the complete input and output features for the baseline MLP models. In the inputs, we include the current state variables like temperature, relative humidity, liquid and ice cloud mixing ratio, zonal and meridional wind. In addition, we also include in the input features the atmosphere gas components (ozone, N2O, and CH4 mixing ratios), boundary conditions (surface pressure, solar insolation and cosine zenith angle, surface heat flux and surface stress, albedo, snow depth), and geography information (land fraction, sea ice fraction). In the output, we predict the time tendencies of temperature, specific humidity, liquid and ice cloud mixing ratio, zonal wind and meridional wind. In addition to the state tendencies, we also predict radiative fluxes and precipitation (snow and rain) at the surface which are required to couple a land model.

Table \ref{additional_input_table} lists the additional input features used in the U-Net-expanded and U-Net-expanded-constrained models. These additional inputs include large-scale forcing at the current and previous time steps, and the convective memory (i.e., the target tendencies) at two previous time steps. Large-scale forcing and convective memory are also used in \cite{han2020moist, wang2022stable, han2023ensemble}. Cosine and sine of latitudes are included as well. When microphysical constraints are added, mixing ratios and previous steps' tendencies of liquid cloud and ice cloud inputs are replaced with the corresponding total cloud condensate inputs (liquid plus ice), along with a diagnosed fraction of liquid cloud based on temperature. That is, the last three rows in the Table \ref{additional_input_table} are only used in the U-Net-expanded-constrained model, which also excludes $q_l(z)$, $q_i(z)$, $dq_l(z,t_{-1},t_{-2})$, and $dq_i(z,t_{-1},t_{-2})$ from the inputs.

\subsection{NN architecture}
\label{appendix:architecture}
\textbf{MLP:} For the results presented in the main text, We used the architecture parameters recommended from the hyperparameter search in \cite{yu2024climsim} in our MLP model. It contains 3 hidden layers with \textit{N}$_\text{nodes} =$ [384, 1024, 640] and ReLU activation. Additionally, we explored an alternative MLP model with \textit{N}$_\text{nodes} =$ [256, 256, 256, 256, 256, 256, 256, 256, 256] and ReLU activation, as proposed in \cite{rasp2018deep}. We observed similar offline and online performance between the two configurations. The results from the second configuration are not shown in the manuscript.


\textbf{U-Net:} We adapted the 2D U-Net model from \cite{song2020score} into a 1D version for column-to-column prediction. The architecture schematic is shown in Figure \ref{U-Net}. Each ResBlock consists of the following operations:
\begin{align*}
y & = \mathrm{Conv1D}(\mathrm{GM}(\mathrm{Conv1D}(\mathrm{silu}(\mathrm{GM}(x))))) + x
\end{align*}
This series of operations includes group normalization (GM), sigmoid linear unit (silu) activation function, 1D convolution (Conv1D) with a kernel size of 3, and a residual connection. The U-Net has 4 layers in depth with latent feature dimensions in each layer \textit{N}$_\text{latent} =$ [128, 256, 256, 256]. The model comprises 13 million parameters.

To evaluate the inference speed of these architectures, we configured the reference E3SM-MMF simulation and hybrid simulations on the NERSC Perlmutter system \url{https://www.nersc.gov/systems/perlmutter}. The simulations were run on a single GPU node equipped with four A100 GPUs, using 4 MPI tasks and 8 threads per task. The reference E3SM-MMF simulation achieves a throughput of 9.9 simulation years per day (SYPD), leveraging GPU acceleration for cloud-resolving model calculations. In comparison, the hybrid MLP simulation achieves a slightly higher throughput of 10.1 SYPD, while the U-Net-baseline simulation achieves 5.2 SYPD. Notably, the neural network computations in these hybrid simulations were performed entirely on CPUs due to unresolved engineering challenges, and GPU acceleration was not utilized in our hybrid simulations. We anticipate that enabling NN inference on GPUs could significantly enhance the throughput of hybrid simulations, making them more computationally efficient than the reference E3SM-MMF simulations. It is important to note that these timing metrics are not very meaningful as our work is not concerned with unthrottling the maximum throughput of hybrid simulations, but rather proving what it takes to encourage their success from a scientific design perspective. It is logical to expect ML model distillation and hardware acceleration strategies can then recover the high throughput expected of hybridized simulations, which is a separate topic.

\subsection{Normalization}
\label{appendix:normalization}
The normalization methods for each input and output feature are listed in Table \ref{input_table}, \ref{output_table}, and \ref{additional_input_table}. The majority of input features are normalized by subtracting the mean and then dividing by the difference of maximum and minimum values. All output variables are normalized by the standard deviation. Mean, maximum, minimum, and standard deviation values are all evaluated per level. For variables that are already in the range of 0 to 1, no normalization is performed. For the large-scale forcings, solar insolation, and surface latent/sensible heat fluxes, we choose not to subtract the mean to maintain their sign. Convective memory inputs are normalized by the same standard deviation value used in the corresponding output variable.

For input liquid, ice, and total cloud mixing ratio, an exponential transformation ($x'=1-exp(-\lambda x)$) is taken to deal with the long positive tail in the distribution of cloud mixing ratio. The rationale is that, if the tail follows an exponential distribution, then an exponential transformation is able to convert the exponential distribution into a 0-1 uniform distribution. The exponential parameter $\lambda$ is evaluated by one over the conditional mean cloud mixing ratio for grids with mixing ratio greater than $10^{-7} kg/kg$, i.e., $\lambda = 1/mean(q_n(q_n>10^{-7}))$.

When calculating the standard deviation for the output tendencies of water vapor, total cloud, zonal and meridional winds, an arbitrary lower bound is added and the actual normalization factor is min(st.d., threshold). These thresholds are added to penalize the contribution of stratosphere levels in the total loss function. In the stratosphere, the standard deviation of these tendencies dramatically decreases with height. This threshold is chosen as $3^{-10}$ kg/kg for tendencies of water vapor and cloud and as $10^{-6}$ m/s for zonal and meridional winds. The tendencies at those levels are also likely noisy and less physical given they are within the sponge layer of the CRM which damps the perturbation. This penalizes water vapor above level 22 (200 hPa), clouds above level 17 (70 hPa), and winds above level 15 (50h Pa). Correspondingly, we also set the top 15 levels of NN output tendencies of water vapor, total cloud, zonal and meridional winds to 0.

\subsection{Additional Preprocessing}
\label{appendix:additional_preprocessing}
After normalization, we concatenate all the input features and output targets into two one-dimensional arrays for each sample and feed them to the MLP model. For the U-Net models, we construct the input features for each sample into a two-dimensional array with a vertical length of 60 and a channel length equal to the number of features. Each profile input feature represents an individual channel, while scalar features such as solar insolation or surface heat fluxes are broadcast as channels with constant values. To facilitate three downsampling steps in the U-Net model, we concatenate four additional layers of zeros at the top of the atmosphere, expanding the vertical column length to 64. The direct output of the U-Net model is a sequence of length 64 with 13 channels. We then drop the top 4 levels in the output. The first 5 output channels correspond to the profile output variables, while the last 8 channels are averaged through the level dimension to obtain the 8 scalar output variables.  

\subsection{Feature Pruning}

The top levels in the cloud-resolving model are within the sponge layer, where the tendencies of water vapor, clouds, and winds can be unphysical and unpredictable. Our output normalization approach also tends to penalize the upper-level tendencies in the total loss function. We do not expect our neural networks can well predict those levels. Therefore, we set the top 15 levels to zero for the predicted tendencies of water vapor, liquid and ice clouds, and zonal and meridional winds, excluding them from the loss function. Additionally, we manually prune the liquid, ice, or total cloud mixing ratio in the top 15 levels (above 50 hPa) from the input by setting them to zero before feeding them to the neural network. This pruning decision is somewhat arbitrary, and we did not test its necessity. A more careful normalization of the output tendencies in the stratosphere could potentially improve their prediction without negatively affecting the fitting skill at lower height levels. In such a case, it would be more appropriate to evaluate the necessity of this pruning.

Previous studies, such as \citeA{brenowitz2018prognostic}, suggested that excluding stratosphere temperature and water vapor (e.g., above around 200 hPa) could be crucial for achieving stable online simulations. To test this hypothesis, we also experimented with a more aggressive pruning strategy, pruning water vapor and clouds above 170 hPa, as well as pruning large-scale forcing and convective memories above 50 hPa. However, we did not observe significant differences with or without this aggressive pruning strategy. All the results presented in this manuscript do not employ this aggressive pruning.

\section{Training}
\label{appendix:training}
Previous studies, such as \cite{ott2020,wang2022stable}, have shown that improved offline skill can benefit downstream error, but the downstream error is not fully constrained by offline skill. This means different checkpoints with very similar offline skills can exhibit varying downstream stability and error. A trial-and-error approach is often used to identify the checkpoint with the best downstream performance. For subtle sensitivities, hundreds of trials can become important to detect online signals of hybrid climate error from upstream ML noise \cite{lin2023sampling}. 

To account for some of this downstream uncertainty, we trained three MLP models with different loss functions and learning rate schedules listed below. We trained these MLP models with a batch size of 1024 using the Adam optimizer. The training uses Distributed Data Parallel on 4 GPUs.
\begin{itemize}
  \item The first model used Mean Absolute Error (MAE) loss, with an initial learning rate of \(1 \times 10^{-3}\), reduced by a factor of 0.3162 every 7 epochs for a total of 28 epochs.
  \item The second model used the same learning rate schedule but a standard Huber loss with \(\delta=1\).
  \item The third model also used Huber loss but with a different learning rate schedule: starting at \(1 \times 10^{-3}\), using a ReduceOnPlateau scheduler with a patience of 3 epochs and a reduction factor of 0.3162 for a total of 20 epochs. This was followed by manually reducing the learning rate to \(1 \times 10^{-4}\) for fine-tuning over 12 more epochs with a patience of 0 epochs and a reduction factor of 0.5.
\end{itemize}

Similarly, we trained three versions of the U-Net models for each of the three configurations: without expanded input features (U-Net-baseline), with expanded input features (U-Net-expanded), and with both expanded input features and cloud physics constraints (U-Net-expanded-constrained). We used the same batch size of 1024 and the Adam optimizer. The training uses Distributed Data Parallel on 4 GPUs.
\begin{itemize}
  \item The first used MAE loss, with an initial learning rate of \(1 \times 10^{-4}\), reduced by a factor of 0.5 every 3 epochs for a total of 16 epochs.
  \item The second model used the same learning rate schedule but a standard Huber loss with \(\delta=1\).
  \item The third used Huber loss with a different learning rate schedule: starting at \(1 \times 10^{-4}\) with a ReduceOnPlateau scheduler with a patience of 3 epochs and a reduction factor of 0.3162 for a total of 20 epochs. For U-Net-expanded and U-Net-expanded-constrained, the learning rate does not drop within the first 20 epochs. Therefore, for these two configurations, we further manually reduced the learning rate to \(5 \times 10^{-5}\) for fine-tuning over 8 more epochs with a patience of 0 epochs and a reduction factor of 0.5.
\end{itemize}

\section{Online Metric}
\label{appendix:online_metric}

In this section, we document how the two main online metrics below are calculated.

\textbf{Root Mean Square Error:} The root mean square error (RMSE) are computed separately for each variable in the online simulations. The goal is to measure the error in simulated climate by analyzing state variables that are sufficiently averaged in space and time. For a given evaluation timescale, RMSE for each variable is calculated as follows:

$$\text{RMSE} = \sqrt{\sum\limits_{i=1}^{S_m} w_i (\hat{y}_m - y_m)^2}$$

where:
\begin{itemize}
  \item $S_m$ is the number of samples. Each grid cell counts as one sample. For a variable evaluated at a specific vertical level, $S_m$ includes all grid cells at that level. For global evaluations, $S_m$ includes all grid cells in the 3D domain (horizontal and vertical).
  \item $\hat{y}_m$ represents the fields from the ML-coupled simulation averaged over the given evaluation timescale. In Section \ref{section:online-1year}, $\hat{y}_m$ represents monthly averaged data. In Section \ref{section:online-5year}, $\hat{y}_m$ represents five-year mean data.
  \item $y_m$ represents the mean values from the reference simulation averaged over the same timescale as $\hat{y}_m$.
  \item $w_1, w_2, \ldots, w_{S_m}$ are mass-weights that sum to 1, proportional to the mass in each grid cell.
\end{itemize}

\textbf{Zonal Mean Bias:} Additionally, we evaluate the long-term zonal mean bias, which measures the average difference between the online simulation and the reference simulation across various atmospheric variables, such as temperature, water vapor, wind, and liquid and ice clouds. The zonal mean bias is derived by comparing variables averaged over time and longitudes. The E3SM-MMF climate model uses unstructured grids instead of regular latitude-longitude grids. To average a variable over all longitudes, we first define latitude bins. For the low-resolution version of the ClimSim dataset, we choose 10-degree intervals from 90°S to 90°N. Within each bin, we count all the grid columns that fall into the bin and calculate the horizontal average of those columns weighted by their area.



\section{Microphysics classifier}

\begin{figure}
\includegraphics[width=1\textwidth]{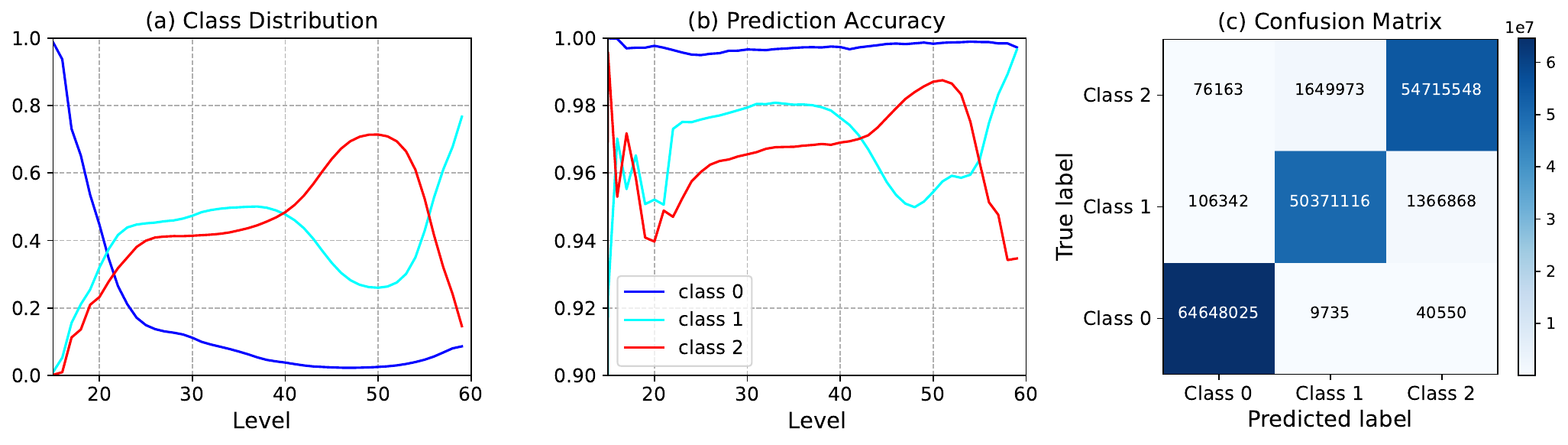}
\centering
\caption{
(a) The distribution of the three classes at each atmospheric level. Only the levels below level 15 are shown since it is mostly Class 0 above level 15. (b) The prediction accuracy of the classifier for each class. Prediction accuracy is defined as the fraction of samples of a given class that are correctly identified by the classifier. (c) The confusion matrix of the classifier for all grid samples at all levels.
}
\label{classifier}
\end{figure}

The tendencies of liquid or ice cloud mixing ratios follow a mixture of discrete and continuous distributions. Some clear-sky grids may have no clouds and, consequently, zero cloud tendencies. However, a naive neural network regression model is likely to predict very small but non-zero values on these grids. In our hybrid model, a non-negative check sets any negative water vapor or cloud mixing ratios to zero. \cite{brenowitz2022emulating} suggested that a separate classifier model can help mitigate the online accumulation of errors in the discrete mode because the non-negative check removes negative errors but retains positive errors. Similarly, \cite{gettelman2021machine} used a classifier as part of their model to emulate a microphysics scheme. Therefore, we trained a microphysics classifier to test if it could help reduce the online cloud bias.

Following \citeA{brenowitz2022emulating}, we defined three classes for cloud changes in the reference E3SM-MMF simulations:

\begin{itemize}
  \item \textbf{Class 0}: No cloud change (defined as the tendency of total cloud mixing ratio smaller than 10$^{-13}$ kg/kg/s).
  \item \textbf{Class 1}: Not Class 0, and all cloud is removed (defined as remaining total cloud mixing ratio smaller than 10$^{-10}$ kg/kg).
  \item \textbf{Class 2}: Not Class 0 or Class 1.
\end{itemize}

Figure \ref{classifier}a shows that a majority of the grids are in either Class 0 or Class 1. Only less than 50\% of grids above level 40 are within Class 2. We trained a classifier model using the same U-Net architecture as before, except with only 1/4 of the latent channel number and 1/2 of the number of Resblocks. The classifier takes the same input features but predicts the logits of each class at each level. We trained the classifier to optimize a cross-entropy loss.

Our classifier model can accurately predict the defined classes but does not significantly change the online error. Figures \ref{classifier}b and \ref{classifier}c show that the classifier is very accurate, with per-level accuracy exceeding 94\% for all classes across all levels. However, when we implemented the classifier into the hybrid E3SM model and masked the regression model's cloud prediction for Class 0 and Class 1 (by setting cloud tendencies to 0 for Class 0 and cloud mixing ratio to 0 for Class 1), we observed little change in the zonal mean climatology with or without the microphysics classifier. We believe that while the classifier prevents the accumulation of prediction errors in the discrete mode, as suggested by \citeA{brenowitz2022emulating}, this error is small compared to the prediction error in grids of Class 2 in our hybrid simulations. This likely explains why adding the microphysics classifier does not significantly improve online performance. All the results presented outside this section do not use a microphysics classifier.

\section{Online Zonal Mean Climatology Bias with a Different U-Net Checkpoint}
\label{appendix:online-bias-another-checkpoint}

\begin{figure}
\includegraphics[width=1\textwidth]{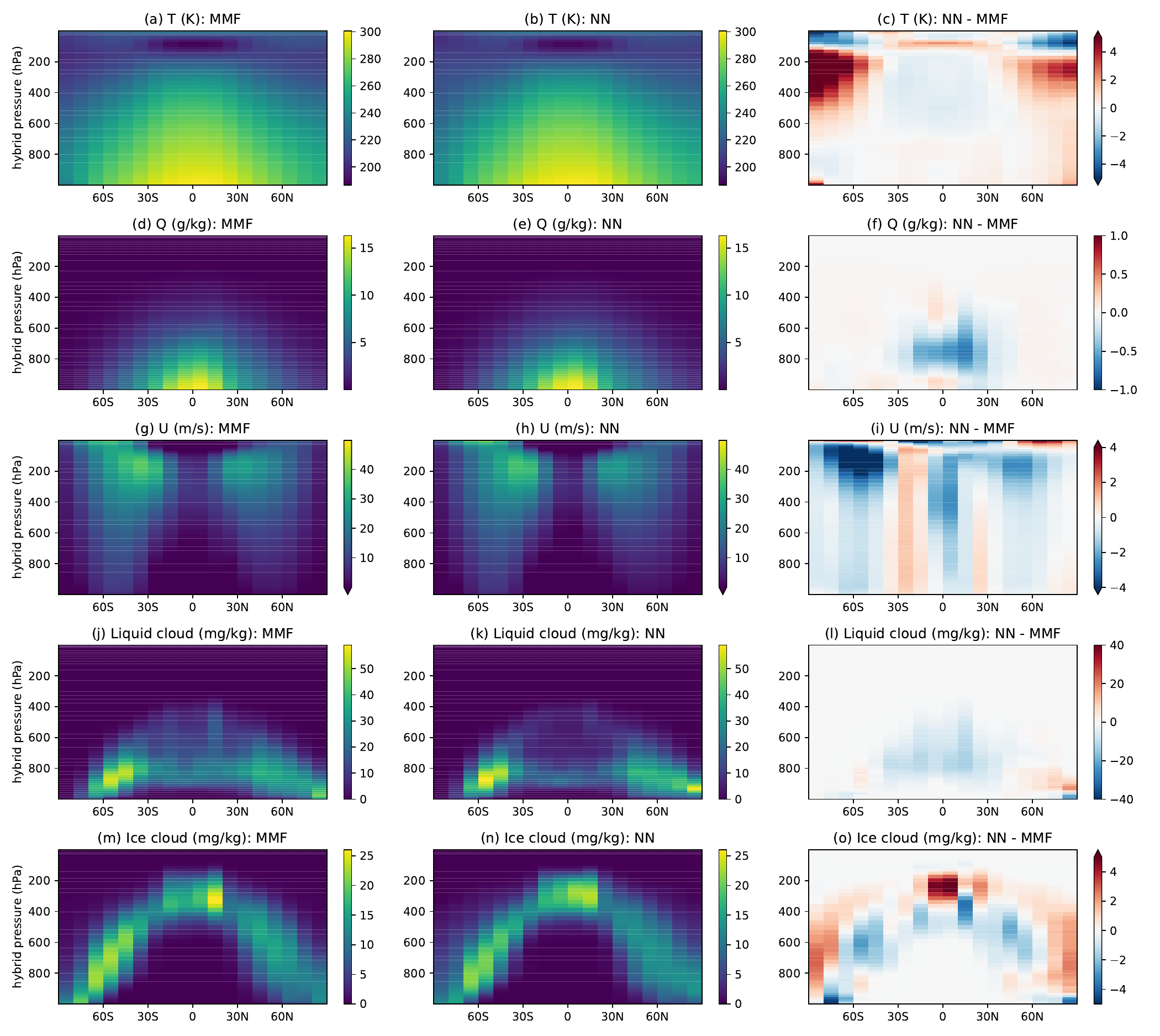}
\centering
\caption{
Similar to Figure \ref{online-states}, but using the U-Net-expanded-constrained model with Huber loss and a different learning rate schedule, as described in \ref{appendix:training}.
}
\label{online-states-another-checkpoint}
\end{figure}

Figure \ref{online-states-another-checkpoint} shows the 5-year zonal-mean climatology using the constrained U-Net model, similar to Figure \ref{online-states}, but with a different U-Net checkpoint due to the use of a different learning rate schedule. This checkpoint exhibits a better extratropical cloud water bias, with only around 10 g/kg cloud water bias compared to the 40 g/kg bias near 60N shown in Figure \ref{online-states}l. When comparing Figure \ref{online-states-another-checkpoint} to Figure \ref{online-states}, many bias patterns remain consistent, such as the vertical dipole temperature bias in the polar stratosphere, a drying water vapor bias and a liquid cloud low bias in the tropical lower troposphere.

\section{Online error growth}
\label{appendix:error-growth}

\begin{figure}
\includegraphics[width=1\textwidth]{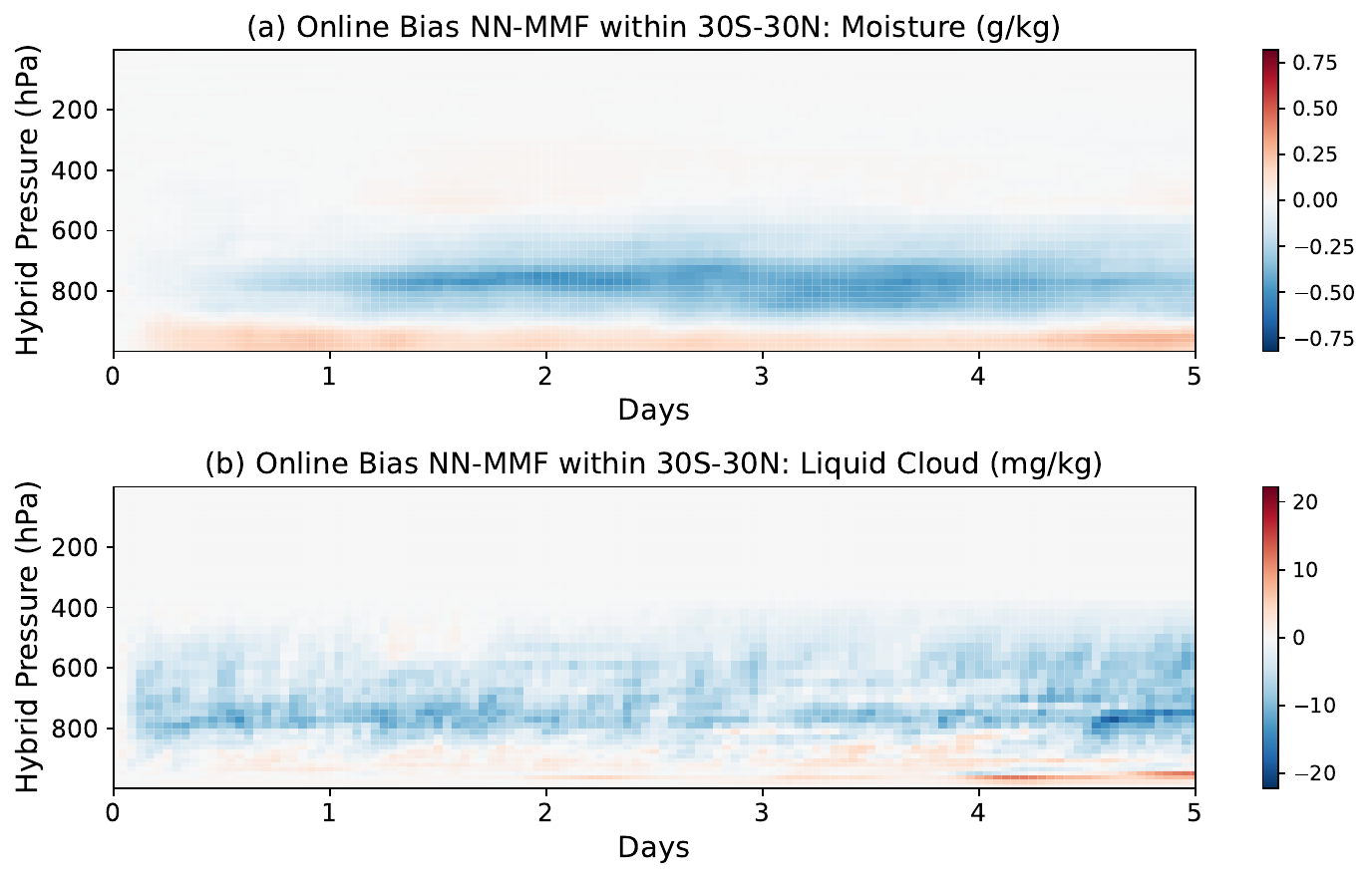}
\centering
\caption{
Time evolution of the zonal mean bias averaged within 30S-30N for water vapor (a) and liquid cloud mixing ratio (b) over the first 5 days in the hybrid simulation with the U-Net model incorporating the microphysical constraints.
}
\label{error-growth}
\end{figure}

We evaluate the online error growth of the hybrid simulation with a constrained U-Net model in comparison to the reference E3SM-MMF simulation. At each time step, a zonal mean bias is calculated and then averaged within 30S-30N. Figure \ref{error-growth} visualizes the tropical mean zonal mean bias for the first 5 days using a pressure-versus-time method. The tropical lower troposphere drying bias develops within the first day and is comparable to the 5-year zonal mean bias shown in Figure \ref{online-states}f. The tropical lower troposphere bias of reduced liquid cloud develops even faster, within only a few hours, and becomes comparable to the 5-year zonal mean bias shown in Figure \ref{online-states}l. This suggests that online biases in the first few simulation days can be indicative of yearly mean online biases. Better understanding of these rapidly developing bias patterns could potentially help in optimizing them.


\acknowledgments
We thank the three anonymous reviewers and the associate editor for their helpful and constructive comments. We thank Pierre Gentine, Tom Buecler, Gunnar Behrens, Thorsten Kurth, Chenggong Wang, Piyush Garg, Laura Zanna, Dale Durran, Yair Cohen, Suman Ravuri for helpful discussion and comments. We thank the NVIDIA Modulus team for the modulus software support. ZK was supported by NASA Grant 80NSSC22K1837, which also supported ZH during the final phase of the project. We thank the computing resources at the NERSC Perlmutter cluster. JL was supported by the National Science Foundation (NSF) Science and Technology Center (STC) Learning the Earth with Artificial Intelligence and Physics (LEAP), Award \# 2019625-STC.
This work was performed under the auspices of the U.S. Department of Energy by Lawrence Livermore National Laboratory under Contract DE-AC52-07NA27344.
WH was supported by the Exascale Computing Project (17-SC-20-SC), a collaborative effort of the U.S. Department of Energy Office of Science and the National Nuclear Security Administration and by the Energy Exascale Earth System Model (E3SM) project, funded by the U.S. Department of Energy, Office of Science, Office of Biological and Environmental Research.



%
\bibliography{citation}

\begin{thebibliography}{}

\bibitem [\protect \citeauthoryear {%
Alexeev%
}{%
Alexeev%
}{%
{\protect \APACyear {2023}}%
}]{%
torchfort}
\APACinsertmetastar {%
torchfort}%
\begin{APACrefauthors}%
Alexeev, D.%
\end{APACrefauthors}%
\unskip\
\newblock
\APACrefYearMonthDay{2023}{{\APACmonth{04}}}{}.
\newblock
\APACrefbtitle {alexeedm/pytorch-fortran: Version v0.4 [Software].} {alexeedm/pytorch-fortran: Version v0.4 [software].}
\newblock
\APACaddressPublisher{}{Zenodo}.
\newblock
\begin{APACrefURL} \url{https://doi.org/10.5281/zenodo.7851167} \end{APACrefURL}
\newblock
\begin{APACrefDOI} \doi{10.5281/zenodo.7851167} \end{APACrefDOI}
\PrintBackRefs{\CurrentBib}

\bibitem [\protect \citeauthoryear {%
Behrens%
\ \protect \BOthers {.}}{%
Behrens%
\ \protect \BOthers {.}}{%
{\protect \APACyear {2024}}%
}]{%
behrens2024improving}
\APACinsertmetastar {%
behrens2024improving}%
\begin{APACrefauthors}%
Behrens, G.%
, Beucler, T.%
, Iglesias-Suarez, F.%
, Yu, S.%
, Gentine, P.%
, Pritchard, M.%
\BDBL {}Eyring, V.%
\end{APACrefauthors}%
\unskip\
\newblock
\APACrefYearMonthDay{2024}{}{}.
\newblock
{\BBOQ}\APACrefatitle {Improving Atmospheric Processes in Earth System Models with Deep Learning Ensembles and Stochastic Parameterizations} {Improving atmospheric processes in earth system models with deep learning ensembles and stochastic parameterizations}.{\BBCQ}
\newblock
\APACjournalVolNumPages{arXiv preprint arXiv:2402.03079}{}{}{}.
\PrintBackRefs{\CurrentBib}

\bibitem [\protect \citeauthoryear {%
Benedict%
\ \BBA {} Randall%
}{%
Benedict%
\ \BBA {} Randall%
}{%
{\protect \APACyear {2009}}%
}]{%
Benedict2009}
\APACinsertmetastar {%
Benedict2009}%
\begin{APACrefauthors}%
Benedict, J\BPBI J.%
\BCBT {}\ \BBA {} Randall, D\BPBI A.%
\end{APACrefauthors}%
\unskip\
\newblock
\APACrefYearMonthDay{2009}{}{}.
\newblock
{\BBOQ}\APACrefatitle {Structure of the Madden--Julian oscillation in the superparameterized CAM} {Structure of the madden--julian oscillation in the superparameterized cam}.{\BBCQ}
\newblock
\APACjournalVolNumPages{J. Atmos. Sci.}{66}{11}{3277--3296}.
\PrintBackRefs{\CurrentBib}

\bibitem [\protect \citeauthoryear {%
Beucler%
\ \protect \BOthers {.}}{%
Beucler%
\ \protect \BOthers {.}}{%
{\protect \APACyear {2024}}%
}]{%
beucler2024climate}
\APACinsertmetastar {%
beucler2024climate}%
\begin{APACrefauthors}%
Beucler, T.%
, Gentine, P.%
, Yuval, J.%
, Gupta, A.%
, Peng, L.%
, Lin, J.%
\BDBL {}others%
\end{APACrefauthors}%
\unskip\
\newblock
\APACrefYearMonthDay{2024}{}{}.
\newblock
{\BBOQ}\APACrefatitle {Climate-invariant machine learning} {Climate-invariant machine learning}.{\BBCQ}
\newblock
\APACjournalVolNumPages{Science Advances}{10}{6}{eadj7250}.
\PrintBackRefs{\CurrentBib}

\bibitem [\protect \citeauthoryear {%
Bhouri%
, Peng%
, Pritchard%
\BCBL {}\ \BBA {} Gentine%
}{%
Bhouri%
\ \protect \BOthers {.}}{%
{\protect \APACyear {2023}}%
}]{%
bhouri2023multi}
\APACinsertmetastar {%
bhouri2023multi}%
\begin{APACrefauthors}%
Bhouri, M\BPBI A.%
, Peng, L.%
, Pritchard, M\BPBI S.%
\BCBL {}\ \BBA {} Gentine, P.%
\end{APACrefauthors}%
\unskip\
\newblock
\APACrefYearMonthDay{2023}{}{}.
\newblock
{\BBOQ}\APACrefatitle {Multi-fidelity climate model parameterization for better generalization and extrapolation} {Multi-fidelity climate model parameterization for better generalization and extrapolation}.{\BBCQ}
\newblock
\APACjournalVolNumPages{arXiv preprint arXiv:2309.10231}{}{}{}.
\PrintBackRefs{\CurrentBib}

\bibitem [\protect \citeauthoryear {%
Brenowitz%
, Beucler%
, Pritchard%
\BCBL {}\ \BBA {} Bretherton%
}{%
Brenowitz%
\ \protect \BOthers {.}}{%
{\protect \APACyear {2020}}%
}]{%
brenowitz2020interpreting}
\APACinsertmetastar {%
brenowitz2020interpreting}%
\begin{APACrefauthors}%
Brenowitz, N\BPBI D.%
, Beucler, T.%
, Pritchard, M.%
\BCBL {}\ \BBA {} Bretherton, C\BPBI S.%
\end{APACrefauthors}%
\unskip\
\newblock
\APACrefYearMonthDay{2020}{}{}.
\newblock
{\BBOQ}\APACrefatitle {Interpreting and stabilizing machine-learning parametrizations of convection} {Interpreting and stabilizing machine-learning parametrizations of convection}.{\BBCQ}
\newblock
\APACjournalVolNumPages{Journal of the Atmospheric Sciences}{77}{12}{4357--4375}.
\PrintBackRefs{\CurrentBib}

\bibitem [\protect \citeauthoryear {%
Brenowitz%
\ \BBA {} Bretherton%
}{%
Brenowitz%
\ \BBA {} Bretherton%
}{%
{\protect \APACyear {2018}}%
}]{%
brenowitz2018prognostic}
\APACinsertmetastar {%
brenowitz2018prognostic}%
\begin{APACrefauthors}%
Brenowitz, N\BPBI D.%
\BCBT {}\ \BBA {} Bretherton, C\BPBI S.%
\end{APACrefauthors}%
\unskip\
\newblock
\APACrefYearMonthDay{2018}{}{}.
\newblock
{\BBOQ}\APACrefatitle {Prognostic validation of a neural network unified physics parameterization} {Prognostic validation of a neural network unified physics parameterization}.{\BBCQ}
\newblock
\APACjournalVolNumPages{Geophysical Research Letters}{45}{12}{6289--6298}.
\PrintBackRefs{\CurrentBib}

\bibitem [\protect \citeauthoryear {%
Brenowitz%
\ \BBA {} Bretherton%
}{%
Brenowitz%
\ \BBA {} Bretherton%
}{%
{\protect \APACyear {2019}}%
}]{%
brenowitz2019spatially}
\APACinsertmetastar {%
brenowitz2019spatially}%
\begin{APACrefauthors}%
Brenowitz, N\BPBI D.%
\BCBT {}\ \BBA {} Bretherton, C\BPBI S.%
\end{APACrefauthors}%
\unskip\
\newblock
\APACrefYearMonthDay{2019}{}{}.
\newblock
{\BBOQ}\APACrefatitle {Spatially extended tests of a neural network parametrization trained by coarse-graining} {Spatially extended tests of a neural network parametrization trained by coarse-graining}.{\BBCQ}
\newblock
\APACjournalVolNumPages{Journal of Advances in Modeling Earth Systems}{11}{8}{2728--2744}.
\PrintBackRefs{\CurrentBib}

\bibitem [\protect \citeauthoryear {%
Brenowitz%
\ \protect \BOthers {.}}{%
Brenowitz%
\ \protect \BOthers {.}}{%
{\protect \APACyear {2022}}%
}]{%
brenowitz2022emulating}
\APACinsertmetastar {%
brenowitz2022emulating}%
\begin{APACrefauthors}%
Brenowitz, N\BPBI D.%
, Perkins, W\BPBI A.%
, Nugent, J\BPBI M.%
, Watt-Meyer, O.%
, Clark, S\BPBI K.%
, Kwa, A.%
\BDBL {}Bretherton, C\BPBI S.%
\end{APACrefauthors}%
\unskip\
\newblock
\APACrefYearMonthDay{2022}{}{}.
\newblock
{\BBOQ}\APACrefatitle {Emulating Fast Processes in Climate Models} {Emulating fast processes in climate models}.{\BBCQ}
\newblock
\APACjournalVolNumPages{arXiv preprint arXiv:2211.10774}{}{}{}.
\PrintBackRefs{\CurrentBib}

\bibitem [\protect \citeauthoryear {%
Christopoulos%
\ \protect \BOthers {.}}{%
Christopoulos%
\ \protect \BOthers {.}}{%
{\protect \APACyear {2024}}%
}]{%
christopoulos2024online}
\APACinsertmetastar {%
christopoulos2024online}%
\begin{APACrefauthors}%
Christopoulos, C.%
, Lopez-Gomez, I.%
, Beucler, T.%
, Cohen, Y.%
, Kawczynski, C.%
, Dunbar, O\BPBI R.%
\BCBL {}\ \BBA {} Schneider, T.%
\end{APACrefauthors}%
\unskip\
\newblock
\APACrefYearMonthDay{2024}{}{}.
\newblock
{\BBOQ}\APACrefatitle {Online learning of entrainment closures in a hybrid machine learning parameterization} {Online learning of entrainment closures in a hybrid machine learning parameterization}.{\BBCQ}
\newblock
\APACjournalVolNumPages{Journal of Advances in Modeling Earth Systems}{16}{11}{e2024MS004485}.
\PrintBackRefs{\CurrentBib}

\bibitem [\protect \citeauthoryear {%
{ClimSim Team}%
}{%
{ClimSim Team}%
}{%
{\protect \APACyear {{\protect \bibnodate {}}}}%
}]{%
climsimrepo}
\APACinsertmetastar {%
climsimrepo}%
\begin{APACrefauthors}%
{ClimSim Team}.%
\end{APACrefauthors}%
\unskip\
\newblock
\APACrefYearMonthDay{{\protect \bibnodate {}}}{}{}.
\newblock
\APACrefbtitle {ClimSim official repository [Software].} {Climsim official repository [software].}
\newblock
\begin{APACrefDOI} \doi{https://github.com/leap-stc/ClimSim} \end{APACrefDOI}
\PrintBackRefs{\CurrentBib}

\bibitem [\protect \citeauthoryear {%
{E3SM Project}%
}{%
{E3SM Project}%
}{%
{\protect \APACyear {2023}}%
}]{%
e3sm-model}
\APACinsertmetastar {%
e3sm-model}%
\begin{APACrefauthors}%
{E3SM Project}.%
\end{APACrefauthors}%
\unskip\
\newblock
\APACrefYearMonthDay{2023}{{\APACmonth{01}}}{}.
\newblock
\APACrefbtitle {{Energy Exascale Earth System Model (E3SM)}.} {{Energy Exascale Earth System Model (E3SM)}.}
\newblock
\APAChowpublished {[Computer Software] \url{https://dx.doi.org/10.11578/E3SM/dc.20230110.5}}.
\newblock
\begin{APACrefURL} \url{https://dx.doi.org/10.11578/E3SM/dc.20230110.5} \end{APACrefURL}
\newblock
\begin{APACrefDOI} \doi{10.11578/E3SM/dc.20230110.5} \end{APACrefDOI}
\PrintBackRefs{\CurrentBib}

\bibitem [\protect \citeauthoryear {%
Eyring%
\ \protect \BOthers {.}}{%
Eyring%
\ \protect \BOthers {.}}{%
{\protect \APACyear {2016}}%
}]{%
eyring2016overview}
\APACinsertmetastar {%
eyring2016overview}%
\begin{APACrefauthors}%
Eyring, V.%
, Bony, S.%
, Meehl, G\BPBI A.%
, Senior, C\BPBI A.%
, Stevens, B.%
, Stouffer, R\BPBI J.%
\BCBL {}\ \BBA {} Taylor, K\BPBI E.%
\end{APACrefauthors}%
\unskip\
\newblock
\APACrefYearMonthDay{2016}{}{}.
\newblock
{\BBOQ}\APACrefatitle {Overview of the Coupled Model Intercomparison Project Phase 6 (CMIP6) experimental design and organization} {Overview of the coupled model intercomparison project phase 6 (cmip6) experimental design and organization}.{\BBCQ}
\newblock
\APACjournalVolNumPages{Geoscientific Model Development}{9}{5}{1937--1958}.
\PrintBackRefs{\CurrentBib}

\bibitem [\protect \citeauthoryear {%
Gentine%
, Pritchard%
, Rasp%
, Reinaudi%
\BCBL {}\ \BBA {} Yacalis%
}{%
Gentine%
\ \protect \BOthers {.}}{%
{\protect \APACyear {2018}}%
}]{%
gentine2018could}
\APACinsertmetastar {%
gentine2018could}%
\begin{APACrefauthors}%
Gentine, P.%
, Pritchard, M.%
, Rasp, S.%
, Reinaudi, G.%
\BCBL {}\ \BBA {} Yacalis, G.%
\end{APACrefauthors}%
\unskip\
\newblock
\APACrefYearMonthDay{2018}{}{}.
\newblock
{\BBOQ}\APACrefatitle {Could machine learning break the convection parameterization deadlock?} {Could machine learning break the convection parameterization deadlock?}{\BBCQ}
\newblock
\APACjournalVolNumPages{Geophysical Research Letters}{45}{11}{5742--5751}.
\PrintBackRefs{\CurrentBib}

\bibitem [\protect \citeauthoryear {%
Gettelman%
\ \protect \BOthers {.}}{%
Gettelman%
\ \protect \BOthers {.}}{%
{\protect \APACyear {2021}}%
}]{%
gettelman2021machine}
\APACinsertmetastar {%
gettelman2021machine}%
\begin{APACrefauthors}%
Gettelman, A.%
, Gagne, D\BPBI J.%
, Chen, C\BHBI C.%
, Christensen, M.%
, Lebo, Z.%
, Morrison, H.%
\BCBL {}\ \BBA {} Gantos, G.%
\end{APACrefauthors}%
\unskip\
\newblock
\APACrefYearMonthDay{2021}{}{}.
\newblock
{\BBOQ}\APACrefatitle {Machine learning the warm rain process} {Machine learning the warm rain process}.{\BBCQ}
\newblock
\APACjournalVolNumPages{Journal of Advances in Modeling Earth Systems}{13}{2}{e2020MS002268}.
\PrintBackRefs{\CurrentBib}

\bibitem [\protect \citeauthoryear {%
Grabowski%
\ \BBA {} Smolarkiewicz%
}{%
Grabowski%
\ \BBA {} Smolarkiewicz%
}{%
{\protect \APACyear {1999}}%
}]{%
Grabowski1999}
\APACinsertmetastar {%
Grabowski1999}%
\begin{APACrefauthors}%
Grabowski, W\BPBI W.%
\BCBT {}\ \BBA {} Smolarkiewicz, P\BPBI K.%
\end{APACrefauthors}%
\unskip\
\newblock
\APACrefYearMonthDay{1999}{}{}.
\newblock
{\BBOQ}\APACrefatitle {CRCP: A cloud resolving convection parameterization for modeling the tropical convecting atmosphere} {Crcp: A cloud resolving convection parameterization for modeling the tropical convecting atmosphere}.{\BBCQ}
\newblock
\APACjournalVolNumPages{Phys. D: Nonlinear Phenom.}{133}{1-4}{171--178}.
\PrintBackRefs{\CurrentBib}

\bibitem [\protect \citeauthoryear {%
Gustafson%
, Berg%
, Easter%
\BCBL {}\ \BBA {} Ghan%
}{%
Gustafson%
\ \protect \BOthers {.}}{%
{\protect \APACyear {2008}}%
}]{%
gustafson2008explicit}
\APACinsertmetastar {%
gustafson2008explicit}%
\begin{APACrefauthors}%
Gustafson, W\BPBI I.%
, Berg, L\BPBI K.%
, Easter, R\BPBI C.%
\BCBL {}\ \BBA {} Ghan, S\BPBI J.%
\end{APACrefauthors}%
\unskip\
\newblock
\APACrefYearMonthDay{2008}{}{}.
\newblock
{\BBOQ}\APACrefatitle {The Explicit-Cloud Parameterized-Pollutant hybrid approach for aerosol--cloud interactions in multiscale modeling framework models: tracer transport results} {The explicit-cloud parameterized-pollutant hybrid approach for aerosol--cloud interactions in multiscale modeling framework models: tracer transport results}.{\BBCQ}
\newblock
\APACjournalVolNumPages{Environmental Research Letters}{3}{2}{025005}.
\PrintBackRefs{\CurrentBib}

\bibitem [\protect \citeauthoryear {%
Han%
, Zhang%
, Huang%
\BCBL {}\ \BBA {} Wang%
}{%
Han%
\ \protect \BOthers {.}}{%
{\protect \APACyear {2020}}%
}]{%
han2020moist}
\APACinsertmetastar {%
han2020moist}%
\begin{APACrefauthors}%
Han, Y.%
, Zhang, G\BPBI J.%
, Huang, X.%
\BCBL {}\ \BBA {} Wang, Y.%
\end{APACrefauthors}%
\unskip\
\newblock
\APACrefYearMonthDay{2020}{}{}.
\newblock
{\BBOQ}\APACrefatitle {A moist physics parameterization based on deep learning} {A moist physics parameterization based on deep learning}.{\BBCQ}
\newblock
\APACjournalVolNumPages{Journal of Advances in Modeling Earth Systems}{12}{9}{e2020MS002076}.
\PrintBackRefs{\CurrentBib}

\bibitem [\protect \citeauthoryear {%
Han%
, Zhang%
\BCBL {}\ \BBA {} Wang%
}{%
Han%
\ \protect \BOthers {.}}{%
{\protect \APACyear {2023}}%
}]{%
han2023ensemble}
\APACinsertmetastar {%
han2023ensemble}%
\begin{APACrefauthors}%
Han, Y.%
, Zhang, G\BPBI J.%
\BCBL {}\ \BBA {} Wang, Y.%
\end{APACrefauthors}%
\unskip\
\newblock
\APACrefYearMonthDay{2023}{}{}.
\newblock
{\BBOQ}\APACrefatitle {An ensemble of neural networks for moist physics processes, its generalizability and stable integration} {An ensemble of neural networks for moist physics processes, its generalizability and stable integration}.{\BBCQ}
\newblock
\APACjournalVolNumPages{Journal of Advances in Modeling Earth Systems}{15}{10}{e2022MS003508}.
\PrintBackRefs{\CurrentBib}

\bibitem [\protect \citeauthoryear {%
Hannah%
\ \protect \BOthers {.}}{%
Hannah%
\ \protect \BOthers {.}}{%
{\protect \APACyear {2020}}%
}]{%
hannah2020initial}
\APACinsertmetastar {%
hannah2020initial}%
\begin{APACrefauthors}%
Hannah, W\BPBI M.%
, Jones, C\BPBI R.%
, Hillman, B\BPBI R.%
, Norman, M\BPBI R.%
, Bader, D\BPBI C.%
, Taylor, M\BPBI A.%
\BDBL {}others%
\end{APACrefauthors}%
\unskip\
\newblock
\APACrefYearMonthDay{2020}{}{}.
\newblock
{\BBOQ}\APACrefatitle {Initial results from the super-parameterized E3SM} {Initial results from the super-parameterized e3sm}.{\BBCQ}
\newblock
\APACjournalVolNumPages{Journal of Advances in Modeling Earth Systems}{12}{1}{e2019MS001863}.
\PrintBackRefs{\CurrentBib}

\bibitem [\protect \citeauthoryear {%
Heuer%
, Schwabe%
, Gentine%
, Giorgetta%
\BCBL {}\ \BBA {} Eyring%
}{%
Heuer%
\ \protect \BOthers {.}}{%
{\protect \APACyear {2023}}%
}]{%
heuer2023interpretable}
\APACinsertmetastar {%
heuer2023interpretable}%
\begin{APACrefauthors}%
Heuer, H.%
, Schwabe, M.%
, Gentine, P.%
, Giorgetta, M\BPBI A.%
\BCBL {}\ \BBA {} Eyring, V.%
\end{APACrefauthors}%
\unskip\
\newblock
\APACrefYearMonthDay{2023}{}{}.
\newblock
{\BBOQ}\APACrefatitle {Interpretable multiscale Machine Learning-Based Parameterizations of Convection for ICON} {Interpretable multiscale machine learning-based parameterizations of convection for icon}.{\BBCQ}
\newblock
\APACjournalVolNumPages{arXiv preprint arXiv:2311.03251}{}{}{}.
\PrintBackRefs{\CurrentBib}

\bibitem [\protect \citeauthoryear {%
Hohenegger%
\ \protect \BOthers {.}}{%
Hohenegger%
\ \protect \BOthers {.}}{%
{\protect \APACyear {2023}}%
}]{%
hohenegger2023icon}
\APACinsertmetastar {%
hohenegger2023icon}%
\begin{APACrefauthors}%
Hohenegger, C.%
, Korn, P.%
, Linardakis, L.%
, Redler, R.%
, Schnur, R.%
, Adamidis, P.%
\BDBL {}others%
\end{APACrefauthors}%
\unskip\
\newblock
\APACrefYearMonthDay{2023}{}{}.
\newblock
{\BBOQ}\APACrefatitle {ICON-Sapphire: simulating the components of the Earth system and their interactions at kilometer and subkilometer scales} {Icon-sapphire: simulating the components of the earth system and their interactions at kilometer and subkilometer scales}.{\BBCQ}
\newblock
\APACjournalVolNumPages{Geoscientific Model Development}{16}{2}{779--811}.
\PrintBackRefs{\CurrentBib}

\bibitem [\protect \citeauthoryear {%
Hu%
}{%
Hu%
}{%
{\protect \APACyear {2025}}%
}]{%
zenodo_hu}
\APACinsertmetastar {%
zenodo_hu}%
\begin{APACrefauthors}%
Hu, Z.%
\end{APACrefauthors}%
\unskip\
\newblock
\APACrefYearMonthDay{2025}{}{}.
\newblock
\APACrefbtitle {Stable Machine-Learning Parameterization of Subgrid Processes in a Comprehensive Atmospheric Model Learned From Embedded Convection-Permitting Simulations: Data and Visualization Notebooks [Software].} {Stable machine-learning parameterization of subgrid processes in a comprehensive atmospheric model learned from embedded convection-permitting simulations: Data and visualization notebooks [software].}
\newblock
\APACaddressPublisher{}{Zenodo}.
\newblock
\begin{APACrefDOI} \doi{10.5281/zenodo.12797810} \end{APACrefDOI}
\PrintBackRefs{\CurrentBib}

\bibitem [\protect \citeauthoryear {%
Iglesias-Suarez%
\ \protect \BOthers {.}}{%
Iglesias-Suarez%
\ \protect \BOthers {.}}{%
{\protect \APACyear {2024}}%
}]{%
iglesias2024causally}
\APACinsertmetastar {%
iglesias2024causally}%
\begin{APACrefauthors}%
Iglesias-Suarez, F.%
, Gentine, P.%
, Solino-Fernandez, B.%
, Beucler, T.%
, Pritchard, M.%
, Runge, J.%
\BCBL {}\ \BBA {} Eyring, V.%
\end{APACrefauthors}%
\unskip\
\newblock
\APACrefYearMonthDay{2024}{}{}.
\newblock
{\BBOQ}\APACrefatitle {Causally-informed deep learning to improve climate models and projections} {Causally-informed deep learning to improve climate models and projections}.{\BBCQ}
\newblock
\APACjournalVolNumPages{Journal of Geophysical Research: Atmospheres}{129}{4}{e2023JD039202}.
\PrintBackRefs{\CurrentBib}

\bibitem [\protect \citeauthoryear {%
IPCC%
}{%
IPCC%
}{%
{\protect \APACyear {2021}}%
}]{%
IPCC2021}
\APACinsertmetastar {%
IPCC2021}%
\begin{APACrefauthors}%
IPCC.%
\end{APACrefauthors}%
\unskip\
\newblock
\APACrefYear{2021}.
\newblock
\APACrefbtitle {Climate Change 2021: The Physical Science Basis. Contribution of Working Group I to the Sixth Assessment Report of the Intergovernmental Panel on Climate Change} {Climate change 2021: The physical science basis. contribution of working group i to the sixth assessment report of the intergovernmental panel on climate change}.
\PrintBackRefs{\CurrentBib}

\bibitem [\protect \citeauthoryear {%
Kelp%
, Jacob%
, Lin%
\BCBL {}\ \BBA {} Sulprizio%
}{%
Kelp%
\ \protect \BOthers {.}}{%
{\protect \APACyear {2022}}%
}]{%
kelp2022online}
\APACinsertmetastar {%
kelp2022online}%
\begin{APACrefauthors}%
Kelp, M\BPBI M.%
, Jacob, D\BPBI J.%
, Lin, H.%
\BCBL {}\ \BBA {} Sulprizio, M\BPBI P.%
\end{APACrefauthors}%
\unskip\
\newblock
\APACrefYearMonthDay{2022}{}{}.
\newblock
{\BBOQ}\APACrefatitle {An online-learned neural network chemical solver for stable long-term global simulations of atmospheric chemistry} {An online-learned neural network chemical solver for stable long-term global simulations of atmospheric chemistry}.{\BBCQ}
\newblock
\APACjournalVolNumPages{Journal of Advances in Modeling Earth Systems}{14}{6}{e2021MS002926}.
\PrintBackRefs{\CurrentBib}

\bibitem [\protect \citeauthoryear {%
Khairoutdinov%
\ \BBA {} Randall%
}{%
Khairoutdinov%
\ \BBA {} Randall%
}{%
{\protect \APACyear {2001}}%
}]{%
khairoutdinov2001cloud}
\APACinsertmetastar {%
khairoutdinov2001cloud}%
\begin{APACrefauthors}%
Khairoutdinov, M\BPBI F.%
\BCBT {}\ \BBA {} Randall, D\BPBI A.%
\end{APACrefauthors}%
\unskip\
\newblock
\APACrefYearMonthDay{2001}{}{}.
\newblock
{\BBOQ}\APACrefatitle {A cloud resolving model as a cloud parameterization in the NCAR Community Climate System Model: Preliminary results} {A cloud resolving model as a cloud parameterization in the ncar community climate system model: Preliminary results}.{\BBCQ}
\newblock
\APACjournalVolNumPages{Geophysical Research Letters}{28}{18}{3617--3620}.
\PrintBackRefs{\CurrentBib}

\bibitem [\protect \citeauthoryear {%
Khairoutdinov%
\ \BBA {} Randall%
}{%
Khairoutdinov%
\ \BBA {} Randall%
}{%
{\protect \APACyear {2003}}%
}]{%
khairoutdinov2003cloud}
\APACinsertmetastar {%
khairoutdinov2003cloud}%
\begin{APACrefauthors}%
Khairoutdinov, M\BPBI F.%
\BCBT {}\ \BBA {} Randall, D\BPBI A.%
\end{APACrefauthors}%
\unskip\
\newblock
\APACrefYearMonthDay{2003}{}{}.
\newblock
{\BBOQ}\APACrefatitle {Cloud resolving modeling of the ARM summer 1997 IOP: Model formulation, results, uncertainties, and sensitivities} {Cloud resolving modeling of the arm summer 1997 iop: Model formulation, results, uncertainties, and sensitivities}.{\BBCQ}
\newblock
\APACjournalVolNumPages{Journal of the Atmospheric Sciences}{60}{4}{607--625}.
\PrintBackRefs{\CurrentBib}

\bibitem [\protect \citeauthoryear {%
Kochkov%
\ \protect \BOthers {.}}{%
Kochkov%
\ \protect \BOthers {.}}{%
{\protect \APACyear {2024}}%
}]{%
kochkov2024neural}
\APACinsertmetastar {%
kochkov2024neural}%
\begin{APACrefauthors}%
Kochkov, D.%
, Yuval, J.%
, Langmore, I.%
, Norgaard, P.%
, Smith, J.%
, Mooers, G.%
\BDBL {}others%
\end{APACrefauthors}%
\unskip\
\newblock
\APACrefYearMonthDay{2024}{}{}.
\newblock
{\BBOQ}\APACrefatitle {Neural general circulation models for weather and climate} {Neural general circulation models for weather and climate}.{\BBCQ}
\newblock
\APACjournalVolNumPages{Nature}{}{}{1--7}.
\PrintBackRefs{\CurrentBib}

\bibitem [\protect \citeauthoryear {%
Kuang%
}{%
Kuang%
}{%
{\protect \APACyear {2024}}%
}]{%
kuang2024linear}
\APACinsertmetastar {%
kuang2024linear}%
\begin{APACrefauthors}%
Kuang, Z.%
\end{APACrefauthors}%
\unskip\
\newblock
\APACrefYearMonthDay{2024}{}{}.
\newblock
{\BBOQ}\APACrefatitle {Linear time-invariant models of a large cumulus ensemble} {Linear time-invariant models of a large cumulus ensemble}.{\BBCQ}
\newblock
\APACjournalVolNumPages{Journal of the Atmospheric Sciences}{81}{3}{605--627}.
\PrintBackRefs{\CurrentBib}

\bibitem [\protect \citeauthoryear {%
K{\"u}hbacher%
, Iglesias-Suarez%
, Kilbertus%
\BCBL {}\ \BBA {} Eyring%
}{%
K{\"u}hbacher%
\ \protect \BOthers {.}}{%
{\protect \APACyear {2024}}%
}]{%
kuhbacher2024towards}
\APACinsertmetastar {%
kuhbacher2024towards}%
\begin{APACrefauthors}%
K{\"u}hbacher, B.%
, Iglesias-Suarez, F.%
, Kilbertus, N.%
\BCBL {}\ \BBA {} Eyring, V.%
\end{APACrefauthors}%
\unskip\
\newblock
\APACrefYearMonthDay{2024}{}{}.
\newblock
{\BBOQ}\APACrefatitle {Towards Physically Consistent Deep Learning For Climate Model Parameterizations} {Towards physically consistent deep learning for climate model parameterizations}.{\BBCQ}
\newblock
\APACjournalVolNumPages{arXiv preprint arXiv:2406.03920}{}{}{}.
\PrintBackRefs{\CurrentBib}

\bibitem [\protect \citeauthoryear {%
Lin%
\ \protect \BOthers {.}}{%
Lin%
\ \protect \BOthers {.}}{%
{\protect \APACyear {2024}}%
}]{%
lin2023sampling}
\APACinsertmetastar {%
lin2023sampling}%
\begin{APACrefauthors}%
Lin, J.%
, Yu, S.%
, Peng, L.%
, Beucler, T.%
, Wong-Toi, E.%
, Hu, Z.%
\BDBL {}Pritchard, M.%
\end{APACrefauthors}%
\unskip\
\newblock
\APACrefYearMonthDay{2024}{}{}.
\newblock
{\BBOQ}\APACrefatitle {Sampling hybrid climate simulation at scale to reliably improve machine learning parameterization} {Sampling hybrid climate simulation at scale to reliably improve machine learning parameterization}.{\BBCQ}
\newblock
\APACjournalVolNumPages{arXiv preprint arXiv:2309.16177v2}{}{}{}.
\PrintBackRefs{\CurrentBib}

\bibitem [\protect \citeauthoryear {%
Lopez-Gomez%
\ \protect \BOthers {.}}{%
Lopez-Gomez%
\ \protect \BOthers {.}}{%
{\protect \APACyear {2022}}%
}]{%
lopez2022training}
\APACinsertmetastar {%
lopez2022training}%
\begin{APACrefauthors}%
Lopez-Gomez, I.%
, Christopoulos, C.%
, Langeland~Ervik, H\BPBI L.%
, Dunbar, O\BPBI R.%
, Cohen, Y.%
\BCBL {}\ \BBA {} Schneider, T.%
\end{APACrefauthors}%
\unskip\
\newblock
\APACrefYearMonthDay{2022}{}{}.
\newblock
{\BBOQ}\APACrefatitle {Training physics-based machine-learning parameterizations with gradient-free ensemble Kalman methods} {Training physics-based machine-learning parameterizations with gradient-free ensemble kalman methods}.{\BBCQ}
\newblock
\APACjournalVolNumPages{Journal of Advances in Modeling Earth Systems}{14}{8}{e2022MS003105}.
\PrintBackRefs{\CurrentBib}

\bibitem [\protect \citeauthoryear {%
Mooers%
\ \protect \BOthers {.}}{%
Mooers%
\ \protect \BOthers {.}}{%
{\protect \APACyear {2021}}%
}]{%
mooers2021assessing}
\APACinsertmetastar {%
mooers2021assessing}%
\begin{APACrefauthors}%
Mooers, G.%
, Pritchard, M.%
, Beucler, T.%
, Ott, J.%
, Yacalis, G.%
, Baldi, P.%
\BCBL {}\ \BBA {} Gentine, P.%
\end{APACrefauthors}%
\unskip\
\newblock
\APACrefYearMonthDay{2021}{}{}.
\newblock
{\BBOQ}\APACrefatitle {Assessing the potential of deep learning for emulating cloud superparameterization in climate models with real-geography boundary conditions} {Assessing the potential of deep learning for emulating cloud superparameterization in climate models with real-geography boundary conditions}.{\BBCQ}
\newblock
\APACjournalVolNumPages{Journal of Advances in Modeling Earth Systems}{13}{5}{e2020MS002385}.
\PrintBackRefs{\CurrentBib}

\bibitem [\protect \citeauthoryear {%
Norman%
\ \protect \BOthers {.}}{%
Norman%
\ \protect \BOthers {.}}{%
{\protect \APACyear {2022}}%
}]{%
Norman2022}
\APACinsertmetastar {%
Norman2022}%
\begin{APACrefauthors}%
Norman, M\BPBI R.%
, Bader, D\BPBI C.%
, Eldred, C.%
, Hannah, W\BPBI M.%
, Hillman, B\BPBI R.%
, Jones, C\BPBI R.%
\BDBL {}others%
\end{APACrefauthors}%
\unskip\
\newblock
\APACrefYearMonthDay{2022}{}{}.
\newblock
{\BBOQ}\APACrefatitle {Unprecedented cloud resolution in a GPU-enabled full-physics atmospheric climate simulation on OLCF’s summit supercomputer} {Unprecedented cloud resolution in a gpu-enabled full-physics atmospheric climate simulation on olcf’s summit supercomputer}.{\BBCQ}
\newblock
\APACjournalVolNumPages{Int. J. High Perform. Compu. Appl.}{36}{1}{93--105}.
\PrintBackRefs{\CurrentBib}

\bibitem [\protect \citeauthoryear {%
{NVIDIA Modulus Team}%
}{%
{NVIDIA Modulus Team}%
}{%
{\protect \APACyear {{\protect \bibnodate {}}}}%
}]{%
modulus}
\APACinsertmetastar {%
modulus}%
\begin{APACrefauthors}%
{NVIDIA Modulus Team}.%
\end{APACrefauthors}%
\unskip\
\newblock
\APACrefYearMonthDay{{\protect \bibnodate {}}}{}{}.
\newblock
\APACrefbtitle {NVIDIA Modulus [Software].} {Nvidia modulus [software].}
\newblock
\begin{APACrefDOI} \doi{https://github.com/NVIDIA/modulus/tree/main} \end{APACrefDOI}
\PrintBackRefs{\CurrentBib}

\bibitem [\protect \citeauthoryear {%
Ott%
\ \protect \BOthers {.}}{%
Ott%
\ \protect \BOthers {.}}{%
{\protect \APACyear {2020}}%
}]{%
ott2020}
\APACinsertmetastar {%
ott2020}%
\begin{APACrefauthors}%
Ott, J.%
, Pritchard, M.%
, Best, N.%
, Linstead, E.%
, Curcic, M.%
\BCBL {}\ \BBA {} Baldi, P.%
\end{APACrefauthors}%
\unskip\
\newblock
\APACrefYearMonthDay{2020}{}{}.
\newblock
\APACrefbtitle {A Fortran-Keras Deep Learning Bridge for Scientific Computing.} {A fortran-keras deep learning bridge for scientific computing.}
\newblock
\APACrefnote{arxiv:2004.10652}
\PrintBackRefs{\CurrentBib}

\bibitem [\protect \citeauthoryear {%
Pahlavan%
, Hassanzadeh%
\BCBL {}\ \BBA {} Alexander%
}{%
Pahlavan%
\ \protect \BOthers {.}}{%
{\protect \APACyear {2024}}%
}]{%
pahlavan2024explainable}
\APACinsertmetastar {%
pahlavan2024explainable}%
\begin{APACrefauthors}%
Pahlavan, H\BPBI A.%
, Hassanzadeh, P.%
\BCBL {}\ \BBA {} Alexander, M\BPBI J.%
\end{APACrefauthors}%
\unskip\
\newblock
\APACrefYearMonthDay{2024}{}{}.
\newblock
{\BBOQ}\APACrefatitle {Explainable Offline-Online Training of Neural Networks for Parameterizations: A 1D Gravity Wave-QBO Testbed in the Small-Data Regime} {Explainable offline-online training of neural networks for parameterizations: A 1d gravity wave-qbo testbed in the small-data regime}.{\BBCQ}
\newblock
\APACjournalVolNumPages{Geophysical Research Letters}{51}{2}{e2023GL106324}.
\PrintBackRefs{\CurrentBib}

\bibitem [\protect \citeauthoryear {%
Parishani%
\ \protect \BOthers {.}}{%
Parishani%
\ \protect \BOthers {.}}{%
{\protect \APACyear {2018}}%
}]{%
parishani2018insensitivity}
\APACinsertmetastar {%
parishani2018insensitivity}%
\begin{APACrefauthors}%
Parishani, H.%
, Pritchard, M\BPBI S.%
, Bretherton, C\BPBI S.%
, Terai, C\BPBI R.%
, Wyant, M\BPBI C.%
, Khairoutdinov, M.%
\BCBL {}\ \BBA {} Singh, B.%
\end{APACrefauthors}%
\unskip\
\newblock
\APACrefYearMonthDay{2018}{}{}.
\newblock
{\BBOQ}\APACrefatitle {Insensitivity of the cloud response to surface warming under radical changes to boundary layer turbulence and cloud microphysics: Results from the ultraparameterized CAM} {Insensitivity of the cloud response to surface warming under radical changes to boundary layer turbulence and cloud microphysics: Results from the ultraparameterized cam}.{\BBCQ}
\newblock
\APACjournalVolNumPages{Journal of Advances in Modeling Earth Systems}{10}{12}{3139--3158}.
\PrintBackRefs{\CurrentBib}

\bibitem [\protect \citeauthoryear {%
Paszke%
\ \protect \BOthers {.}}{%
Paszke%
\ \protect \BOthers {.}}{%
{\protect \APACyear {2019}}%
}]{%
paszke2019pytorch}
\APACinsertmetastar {%
paszke2019pytorch}%
\begin{APACrefauthors}%
Paszke, A.%
, Gross, S.%
, Massa, F.%
, Lerer, A.%
, Bradbury, J.%
, Chanan, G.%
\BDBL {}others%
\end{APACrefauthors}%
\unskip\
\newblock
\APACrefYearMonthDay{2019}{}{}.
\newblock
{\BBOQ}\APACrefatitle {Pytorch: An imperative style, high-performance deep learning library} {Pytorch: An imperative style, high-performance deep learning library}.{\BBCQ}
\newblock
\APACjournalVolNumPages{Advances in neural information processing systems}{32}{}{}.
\PrintBackRefs{\CurrentBib}

\bibitem [\protect \citeauthoryear {%
Peng%
\ \protect \BOthers {.}}{%
Peng%
\ \protect \BOthers {.}}{%
{\protect \APACyear {2024}}%
}]{%
peng2024improving}
\APACinsertmetastar {%
peng2024improving}%
\begin{APACrefauthors}%
Peng, L.%
, Blossey, P\BPBI N.%
, Hannah, W\BPBI M.%
, Bretherton, C\BPBI S.%
, Terai, C\BPBI R.%
, Jenney, A\BPBI M.%
\BCBL {}\ \BBA {} Pritchard, M.%
\end{APACrefauthors}%
\unskip\
\newblock
\APACrefYearMonthDay{2024}{}{}.
\newblock
{\BBOQ}\APACrefatitle {Improving stratocumulus cloud amounts in a 200-m resolution multi-scale modeling framework through tuning of its interior physics} {Improving stratocumulus cloud amounts in a 200-m resolution multi-scale modeling framework through tuning of its interior physics}.{\BBCQ}
\newblock
\APACjournalVolNumPages{Journal of Advances in Modeling Earth Systems}{16}{3}{e2023MS003632}.
\PrintBackRefs{\CurrentBib}

\bibitem [\protect \citeauthoryear {%
{PyTorch Contributors}%
}{%
{PyTorch Contributors}%
}{%
{\protect \APACyear {2024}}%
}]{%
torchscript}
\APACinsertmetastar {%
torchscript}%
\begin{APACrefauthors}%
{PyTorch Contributors}.%
\end{APACrefauthors}%
\unskip\
\newblock
\APACrefYearMonthDay{2024}{}{}.
\newblock
\APACrefbtitle {TorchScript Documentation [Software].} {Torchscript documentation [software].}
\newblock
\begin{APACrefURL} \url{https://pytorch.org/docs/stable/jit.html#pytorch-functions-and-modules} \end{APACrefURL}
\PrintBackRefs{\CurrentBib}

\bibitem [\protect \citeauthoryear {%
Randall%
}{%
Randall%
}{%
{\protect \APACyear {2013}}%
}]{%
Randall2013}
\APACinsertmetastar {%
Randall2013}%
\begin{APACrefauthors}%
Randall, D\BPBI A.%
\end{APACrefauthors}%
\unskip\
\newblock
\APACrefYearMonthDay{2013}{}{}.
\newblock
{\BBOQ}\APACrefatitle {Beyond deadlock} {Beyond deadlock}.{\BBCQ}
\newblock
\APACjournalVolNumPages{Geophys. Res. Lett.}{40}{22}{5970--5976}.
\PrintBackRefs{\CurrentBib}

\bibitem [\protect \citeauthoryear {%
Rasp%
}{%
Rasp%
}{%
{\protect \APACyear {2020}}%
}]{%
rasp2020coupled}
\APACinsertmetastar {%
rasp2020coupled}%
\begin{APACrefauthors}%
Rasp, S.%
\end{APACrefauthors}%
\unskip\
\newblock
\APACrefYearMonthDay{2020}{}{}.
\newblock
{\BBOQ}\APACrefatitle {Coupled online learning as a way to tackle instabilities and biases in neural network parameterizations: General algorithms and Lorenz 96 case study (v1. 0)} {Coupled online learning as a way to tackle instabilities and biases in neural network parameterizations: General algorithms and lorenz 96 case study (v1. 0)}.{\BBCQ}
\newblock
\APACjournalVolNumPages{Geoscientific Model Development}{13}{5}{2185--2196}.
\PrintBackRefs{\CurrentBib}

\bibitem [\protect \citeauthoryear {%
Rasp%
, Pritchard%
\BCBL {}\ \BBA {} Gentine%
}{%
Rasp%
\ \protect \BOthers {.}}{%
{\protect \APACyear {2018}}%
}]{%
rasp2018deep}
\APACinsertmetastar {%
rasp2018deep}%
\begin{APACrefauthors}%
Rasp, S.%
, Pritchard, M\BPBI S.%
\BCBL {}\ \BBA {} Gentine, P.%
\end{APACrefauthors}%
\unskip\
\newblock
\APACrefYearMonthDay{2018}{}{}.
\newblock
{\BBOQ}\APACrefatitle {Deep learning to represent subgrid processes in climate models} {Deep learning to represent subgrid processes in climate models}.{\BBCQ}
\newblock
\APACjournalVolNumPages{Proceedings of the National Academy of Sciences}{115}{39}{9684--9689}.
\PrintBackRefs{\CurrentBib}

\bibitem [\protect \citeauthoryear {%
A.~Ross%
, Li%
, Perezhogin%
, Fernandez-Granda%
\BCBL {}\ \BBA {} Zanna%
}{%
A.~Ross%
\ \protect \BOthers {.}}{%
{\protect \APACyear {2023}}%
}]{%
ross2023benchmarking}
\APACinsertmetastar {%
ross2023benchmarking}%
\begin{APACrefauthors}%
Ross, A.%
, Li, Z.%
, Perezhogin, P.%
, Fernandez-Granda, C.%
\BCBL {}\ \BBA {} Zanna, L.%
\end{APACrefauthors}%
\unskip\
\newblock
\APACrefYearMonthDay{2023}{}{}.
\newblock
{\BBOQ}\APACrefatitle {Benchmarking of machine learning ocean subgrid parameterizations in an idealized model} {Benchmarking of machine learning ocean subgrid parameterizations in an idealized model}.{\BBCQ}
\newblock
\APACjournalVolNumPages{Journal of Advances in Modeling Earth Systems}{15}{1}{}.
\PrintBackRefs{\CurrentBib}

\bibitem [\protect \citeauthoryear {%
S.~Ross%
, Gordon%
\BCBL {}\ \BBA {} Bagnell%
}{%
S.~Ross%
\ \protect \BOthers {.}}{%
{\protect \APACyear {2011}}%
}]{%
ross2011reduction}
\APACinsertmetastar {%
ross2011reduction}%
\begin{APACrefauthors}%
Ross, S.%
, Gordon, G.%
\BCBL {}\ \BBA {} Bagnell, D.%
\end{APACrefauthors}%
\unskip\
\newblock
\APACrefYearMonthDay{2011}{}{}.
\newblock
{\BBOQ}\APACrefatitle {A reduction of imitation learning and structured prediction to no-regret online learning} {A reduction of imitation learning and structured prediction to no-regret online learning}.{\BBCQ}
\newblock
\BIn{} \APACrefbtitle {Proceedings of the fourteenth international conference on artificial intelligence and statistics} {Proceedings of the fourteenth international conference on artificial intelligence and statistics}\ (\BPGS\ 627--635).
\PrintBackRefs{\CurrentBib}

\bibitem [\protect \citeauthoryear {%
Sanford%
\ \protect \BOthers {.}}{%
Sanford%
\ \protect \BOthers {.}}{%
{\protect \APACyear {2023}}%
}]{%
sanford2023improving}
\APACinsertmetastar {%
sanford2023improving}%
\begin{APACrefauthors}%
Sanford, C.%
, Kwa, A.%
, Watt-Meyer, O.%
, Clark, S\BPBI K.%
, Brenowitz, N.%
, McGibbon, J.%
\BCBL {}\ \BBA {} Bretherton, C.%
\end{APACrefauthors}%
\unskip\
\newblock
\APACrefYearMonthDay{2023}{}{}.
\newblock
{\BBOQ}\APACrefatitle {Improving the reliability of ML-corrected climate models with novelty detection} {Improving the reliability of ml-corrected climate models with novelty detection}.{\BBCQ}
\newblock
\APACjournalVolNumPages{Journal of Advances in Modeling Earth Systems}{15}{11}{e2023MS003809}.
\PrintBackRefs{\CurrentBib}

\bibitem [\protect \citeauthoryear {%
Shamekh%
, Lamb%
, Huang%
\BCBL {}\ \BBA {} Gentine%
}{%
Shamekh%
\ \protect \BOthers {.}}{%
{\protect \APACyear {2023}}%
}]{%
shamekh2023implicit}
\APACinsertmetastar {%
shamekh2023implicit}%
\begin{APACrefauthors}%
Shamekh, S.%
, Lamb, K\BPBI D.%
, Huang, Y.%
\BCBL {}\ \BBA {} Gentine, P.%
\end{APACrefauthors}%
\unskip\
\newblock
\APACrefYearMonthDay{2023}{}{}.
\newblock
{\BBOQ}\APACrefatitle {Implicit learning of convective organization explains precipitation stochasticity} {Implicit learning of convective organization explains precipitation stochasticity}.{\BBCQ}
\newblock
\APACjournalVolNumPages{Proceedings of the National Academy of Sciences}{120}{20}{e2216158120}.
\PrintBackRefs{\CurrentBib}

\bibitem [\protect \citeauthoryear {%
Sherwood%
\ \protect \BOthers {.}}{%
Sherwood%
\ \protect \BOthers {.}}{%
{\protect \APACyear {2020}}%
}]{%
Sherwood2020}
\APACinsertmetastar {%
Sherwood2020}%
\begin{APACrefauthors}%
Sherwood, S.%
, Webb, M\BPBI J.%
, Annan, J\BPBI D.%
, Armour, K\BPBI C.%
, Forster, P\BPBI M.%
, Hargreaves, J\BPBI C.%
\BDBL {}others%
\end{APACrefauthors}%
\unskip\
\newblock
\APACrefYearMonthDay{2020}{}{}.
\newblock
{\BBOQ}\APACrefatitle {An assessment of Earth's climate sensitivity using multiple lines of evidence} {An assessment of earth's climate sensitivity using multiple lines of evidence}.{\BBCQ}
\newblock
\APACjournalVolNumPages{Rev. Geophys.}{58}{4}{e2019RG000678}.
\PrintBackRefs{\CurrentBib}

\bibitem [\protect \citeauthoryear {%
Song%
\ \protect \BOthers {.}}{%
Song%
\ \protect \BOthers {.}}{%
{\protect \APACyear {2020}}%
}]{%
song2020score}
\APACinsertmetastar {%
song2020score}%
\begin{APACrefauthors}%
Song, Y.%
, Sohl-Dickstein, J.%
, Kingma, D\BPBI P.%
, Kumar, A.%
, Ermon, S.%
\BCBL {}\ \BBA {} Poole, B.%
\end{APACrefauthors}%
\unskip\
\newblock
\APACrefYearMonthDay{2020}{}{}.
\newblock
{\BBOQ}\APACrefatitle {Score-based generative modeling through stochastic differential equations} {Score-based generative modeling through stochastic differential equations}.{\BBCQ}
\newblock
\APACjournalVolNumPages{arXiv preprint arXiv:2011.13456}{}{}{}.
\PrintBackRefs{\CurrentBib}

\bibitem [\protect \citeauthoryear {%
Taylor%
\ \protect \BOthers {.}}{%
Taylor%
\ \protect \BOthers {.}}{%
{\protect \APACyear {2023}}%
}]{%
taylor2023simple}
\APACinsertmetastar {%
taylor2023simple}%
\begin{APACrefauthors}%
Taylor, M.%
, Caldwell, P\BPBI M.%
, Bertagna, L.%
, Clevenger, C.%
, Donahue, A.%
, Foucar, J.%
\BDBL {}others%
\end{APACrefauthors}%
\unskip\
\newblock
\APACrefYearMonthDay{2023}{}{}.
\newblock
{\BBOQ}\APACrefatitle {The Simple Cloud-Resolving E3SM Atmosphere Model Running on the Frontier Exascale System} {The simple cloud-resolving e3sm atmosphere model running on the frontier exascale system}.{\BBCQ}
\newblock
\BIn{} \APACrefbtitle {Proceedings of the International Conference for High Performance Computing, Networking, Storage and Analysis} {Proceedings of the international conference for high performance computing, networking, storage and analysis}\ (\BPGS\ 1--11).
\PrintBackRefs{\CurrentBib}

\bibitem [\protect \citeauthoryear {%
Terai%
, Pritchard%
, Blossey%
\BCBL {}\ \BBA {} Bretherton%
}{%
Terai%
\ \protect \BOthers {.}}{%
{\protect \APACyear {2020}}%
}]{%
terai2020impact}
\APACinsertmetastar {%
terai2020impact}%
\begin{APACrefauthors}%
Terai, C.%
, Pritchard, M.%
, Blossey, P.%
\BCBL {}\ \BBA {} Bretherton, C.%
\end{APACrefauthors}%
\unskip\
\newblock
\APACrefYearMonthDay{2020}{}{}.
\newblock
{\BBOQ}\APACrefatitle {The impact of resolving subkilometer processes on aerosol-cloud interactions of low-level clouds in global model simulations} {The impact of resolving subkilometer processes on aerosol-cloud interactions of low-level clouds in global model simulations}.{\BBCQ}
\newblock
\APACjournalVolNumPages{Journal of Advances in Modeling Earth Systems}{12}{11}{e2020MS002274}.
\PrintBackRefs{\CurrentBib}

\bibitem [\protect \citeauthoryear {%
Tulich%
}{%
Tulich%
}{%
{\protect \APACyear {2015}}%
}]{%
tulich2015strategy}
\APACinsertmetastar {%
tulich2015strategy}%
\begin{APACrefauthors}%
Tulich, S.%
\end{APACrefauthors}%
\unskip\
\newblock
\APACrefYearMonthDay{2015}{}{}.
\newblock
{\BBOQ}\APACrefatitle {A strategy for representing the effects of convective momentum transport in multiscale models: Evaluation using a new superparameterized version of the Weather Research and Forecast model (SP-WRF)} {A strategy for representing the effects of convective momentum transport in multiscale models: Evaluation using a new superparameterized version of the weather research and forecast model (sp-wrf)}.{\BBCQ}
\newblock
\APACjournalVolNumPages{Journal of Advances in Modeling Earth Systems}{7}{2}{938--962}.
\PrintBackRefs{\CurrentBib}

\bibitem [\protect \citeauthoryear {%
P.~Wang%
, Yuval%
\BCBL {}\ \BBA {} O’Gorman%
}{%
P.~Wang%
\ \protect \BOthers {.}}{%
{\protect \APACyear {2022}}%
}]{%
wang2022non}
\APACinsertmetastar {%
wang2022non}%
\begin{APACrefauthors}%
Wang, P.%
, Yuval, J.%
\BCBL {}\ \BBA {} O’Gorman, P\BPBI A.%
\end{APACrefauthors}%
\unskip\
\newblock
\APACrefYearMonthDay{2022}{}{}.
\newblock
{\BBOQ}\APACrefatitle {Non-local parameterization of atmospheric subgrid processes with neural networks} {Non-local parameterization of atmospheric subgrid processes with neural networks}.{\BBCQ}
\newblock
\APACjournalVolNumPages{Journal of Advances in Modeling Earth Systems}{14}{10}{e2022MS002984}.
\PrintBackRefs{\CurrentBib}

\bibitem [\protect \citeauthoryear {%
X.~Wang%
, Han%
, Xue%
, Yang%
\BCBL {}\ \BBA {} Zhang%
}{%
X.~Wang%
\ \protect \BOthers {.}}{%
{\protect \APACyear {2022}}%
}]{%
wang2022stable}
\APACinsertmetastar {%
wang2022stable}%
\begin{APACrefauthors}%
Wang, X.%
, Han, Y.%
, Xue, W.%
, Yang, G.%
\BCBL {}\ \BBA {} Zhang, G\BPBI J.%
\end{APACrefauthors}%
\unskip\
\newblock
\APACrefYearMonthDay{2022}{}{}.
\newblock
{\BBOQ}\APACrefatitle {Stable climate simulations using a realistic general circulation model with neural network parameterizations for atmospheric moist physics and radiation processes} {Stable climate simulations using a realistic general circulation model with neural network parameterizations for atmospheric moist physics and radiation processes}.{\BBCQ}
\newblock
\APACjournalVolNumPages{Geoscientific Model Development}{15}{9}{3923--3940}.
\PrintBackRefs{\CurrentBib}

\bibitem [\protect \citeauthoryear {%
Yang%
, Hannah%
\BCBL {}\ \BBA {} Leung%
}{%
Yang%
\ \protect \BOthers {.}}{%
{\protect \APACyear {2022}}%
}]{%
yang2022convective}
\APACinsertmetastar {%
yang2022convective}%
\begin{APACrefauthors}%
Yang, Q.%
, Hannah, W\BPBI M.%
\BCBL {}\ \BBA {} Leung, L\BPBI R.%
\end{APACrefauthors}%
\unskip\
\newblock
\APACrefYearMonthDay{2022}{}{}.
\newblock
{\BBOQ}\APACrefatitle {Convective momentum transport and its impact on the Madden-Julian oscillation in E3SM-MMF} {Convective momentum transport and its impact on the madden-julian oscillation in e3sm-mmf}.{\BBCQ}
\newblock
\APACjournalVolNumPages{Journal of Advances in Modeling Earth Systems}{14}{11}{e2022MS003206}.
\PrintBackRefs{\CurrentBib}

\bibitem [\protect \citeauthoryear {%
Yu%
, Hannah%
\BCBL {}\ \protect \BOthers {.}}{%
Yu%
, Hannah%
\BCBL {}\ \protect \BOthers {.}}{%
{\protect \APACyear {2024}}%
}]{%
yu2024climsim}
\APACinsertmetastar {%
yu2024climsim}%
\begin{APACrefauthors}%
Yu, S.%
, Hannah, W.%
, Peng, L.%
, Lin, J.%
, Bhouri, M\BPBI A.%
, Gupta, R.%
\BDBL {}others%
\end{APACrefauthors}%
\unskip\
\newblock
\APACrefYearMonthDay{2024}{}{}.
\newblock
{\BBOQ}\APACrefatitle {ClimSim: A large multi-scale dataset for hybrid physics-ML climate emulation} {Climsim: A large multi-scale dataset for hybrid physics-ml climate emulation}.{\BBCQ}
\newblock
\APACjournalVolNumPages{Advances in Neural Information Processing Systems}{36}{}{}.
\PrintBackRefs{\CurrentBib}

\bibitem [\protect \citeauthoryear {%
Yu%
, Hu%
\BCBL {}\ \protect \BOthers {.}}{%
Yu%
, Hu%
\BCBL {}\ \protect \BOthers {.}}{%
{\protect \APACyear {2024}}%
}]{%
yu2024climsimonline}
\APACinsertmetastar {%
yu2024climsimonline}%
\begin{APACrefauthors}%
Yu, S.%
, Hu, Z.%
, Subramaniam, A.%
, Hannah, W.%
, Peng, L.%
, Lin, J.%
\BDBL {}Pritchard, M.%
\end{APACrefauthors}%
\unskip\
\newblock
\APACrefYearMonthDay{2024}{}{}.
\newblock
\APACrefbtitle {ClimSim-Online: A Large Multi-scale Dataset and Framework for Hybrid ML-physics Climate Emulation.} {Climsim-online: A large multi-scale dataset and framework for hybrid ml-physics climate emulation.}
\newblock
\begin{APACrefURL} \url{https://arxiv.org/abs/2306.08754} \end{APACrefURL}
\PrintBackRefs{\CurrentBib}

\bibitem [\protect \citeauthoryear {%
Yuval%
, O'Gorman%
\BCBL {}\ \BBA {} Hill%
}{%
Yuval%
\ \protect \BOthers {.}}{%
{\protect \APACyear {2021}}%
}]{%
yuval2021use}
\APACinsertmetastar {%
yuval2021use}%
\begin{APACrefauthors}%
Yuval, J.%
, O'Gorman, P\BPBI A.%
\BCBL {}\ \BBA {} Hill, C\BPBI N.%
\end{APACrefauthors}%
\unskip\
\newblock
\APACrefYearMonthDay{2021}{}{}.
\newblock
{\BBOQ}\APACrefatitle {Use of neural networks for stable, accurate and physically consistent parameterization of subgrid atmospheric processes with good performance at reduced precision} {Use of neural networks for stable, accurate and physically consistent parameterization of subgrid atmospheric processes with good performance at reduced precision}.{\BBCQ}
\newblock
\APACjournalVolNumPages{Geophysical Research Letters}{48}{6}{e2020GL091363}.
\PrintBackRefs{\CurrentBib}

\bibitem [\protect \citeauthoryear {%
Yuval%
\ \BBA {} O’Gorman%
}{%
Yuval%
\ \BBA {} O’Gorman%
}{%
{\protect \APACyear {2020}}%
}]{%
yuval2020stable}
\APACinsertmetastar {%
yuval2020stable}%
\begin{APACrefauthors}%
Yuval, J.%
\BCBT {}\ \BBA {} O’Gorman, P\BPBI A.%
\end{APACrefauthors}%
\unskip\
\newblock
\APACrefYearMonthDay{2020}{}{}.
\newblock
{\BBOQ}\APACrefatitle {Stable machine-learning parameterization of subgrid processes for climate modeling at a range of resolutions} {Stable machine-learning parameterization of subgrid processes for climate modeling at a range of resolutions}.{\BBCQ}
\newblock
\APACjournalVolNumPages{Nature communications}{11}{1}{3295}.
\PrintBackRefs{\CurrentBib}

\end{thebibliography}

%
%




%
%
%
%
%

\end{document}


%
%


\title{Supporting Information for ``Stable Machine-Learning Parameterization of Subgrid Processes in a Comprehensive Atmospheric Model Learned From Embedded Convection-Permitting Simulations"}

\authors{Zeyuan Hu\affil{1,2}\thanks{Work done during an internship at NVIDIA}, Akshay Subramaniam\affil{1}, Zhiming Kuang\affil{2}, Jerry Lin\affil{3}, Sungduk Yu\affil{4}, Walter M. Hannah\affil{5}, Noah D. Brenowitz\affil{1}, Josh Romero\affil{1}, Michael S. Pritchard\affil{1,3}}

\affiliation{1}{NVIDIA Research}
\affiliation{2}{Harvard University}
\affiliation{3}{University of California at Irvine}
\affiliation{4}{Multimodal Cognitive AI, Intel Labs, Santa Clara, CA 95054, USA}
\affiliation{5}{Lawrence Livermore National Laboratory}

%
%

%

\begin{article}

%
%

\noindent\textbf{Contents of this file}

\begin{enumerate}

\item Figures S1 to S3

\end{enumerate}

\end{article}

\begin{figure}[!htbp]
\centering
\includegraphics[width=1\textwidth]{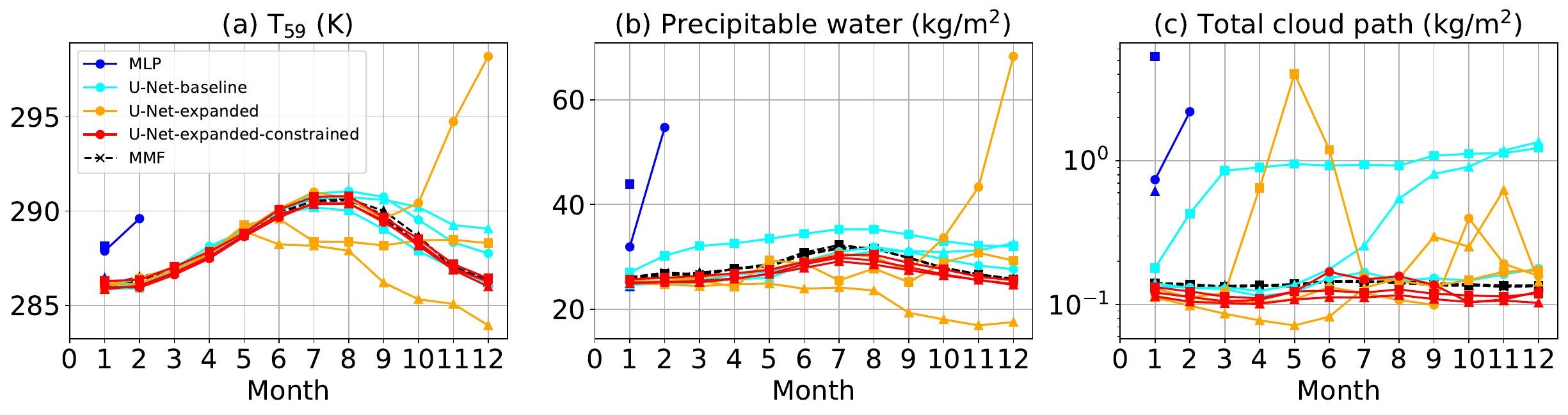}

\caption{Online monthly and globally averaged (a) near-surface temperature at the lowest model level (K), (b) precipitable water (kg/m$^2$), and (c) total (liquid+ice) cloud path (kg/m$^2$) over a one-year period, comparing baseline MLP (blue), U-Net-baseline (cyan), U-Net-expanded (orange), and U-Net-expanded-constrained (red) against the reference E3SM-MMF simulations (dashed black lines). The dashed black lines construct an ensemble of reference physical simulations, and their spread represents internal atmospheric variability. Each architecture was trained with three slightly different configurations (e.g., loss functions and learning rate schedules) to sample ML uncertainty, represented by different markers but with the same line color.}
\label{fig:timeseries-1y-mean}
\end{figure}

\begin{figure}[!htbp]
\centering
\includegraphics[width=1\textwidth]{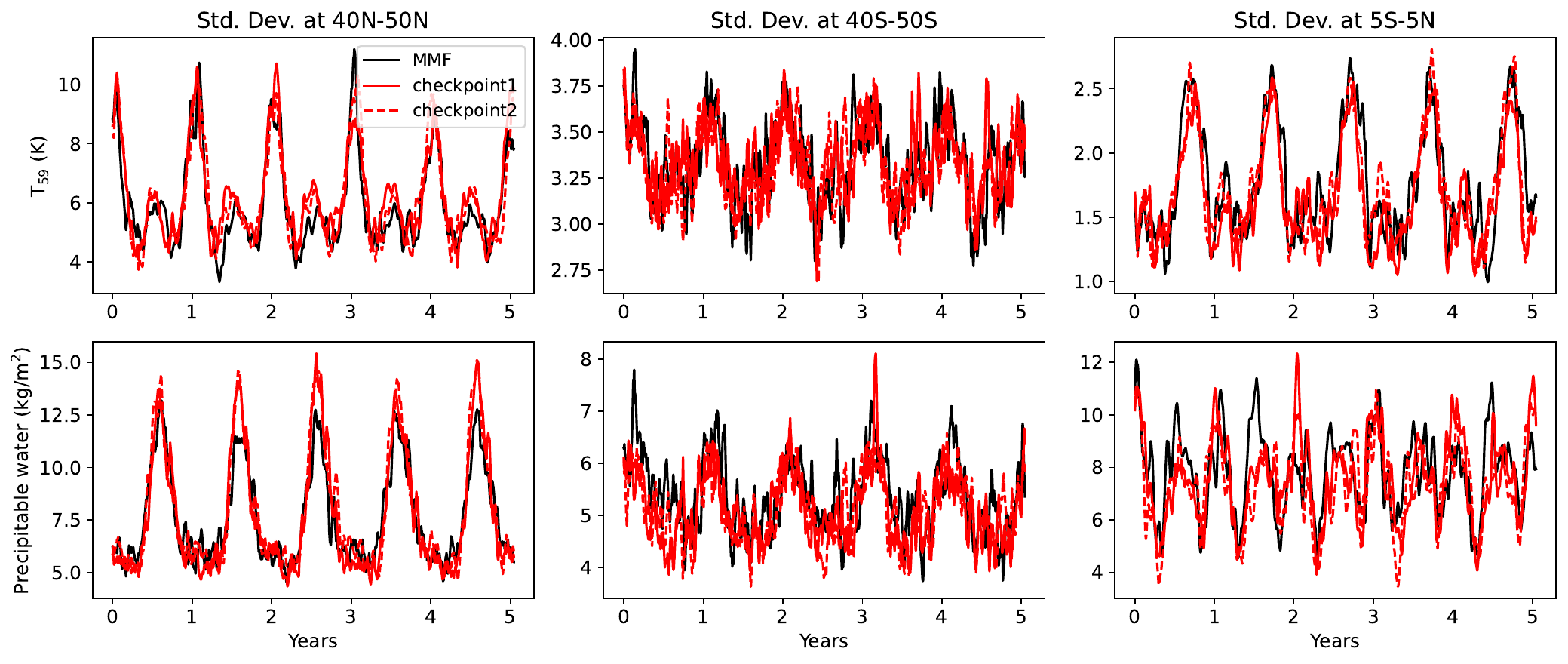}

\caption{Five-year time series of spatial standard deviation of near-surface temperature at the lowest model level (top row) and precipitable water (bottom row). The standard deviation is computed using daily-averaged fields across all grid points within three latitude bands: 40°N–50°N (left column), 40°S–50°S (middle column), and 5°S–5°N (right column). A 15-day running average is applied to smooth the time series. The black line represents the reference E3SM-MMF simulation. The solid and dashed red lines represent the hybrid simulations using two U-Net-expanded-constrained checkpoints: Huber loss with a reduce-on-plateau scheduler (solid red) and Huber loss with a step scheduler (dashed red).}
\label{fig:std}
\end{figure}

\begin{figure}[!htbp]
\centering
\includegraphics[width=1\textwidth]{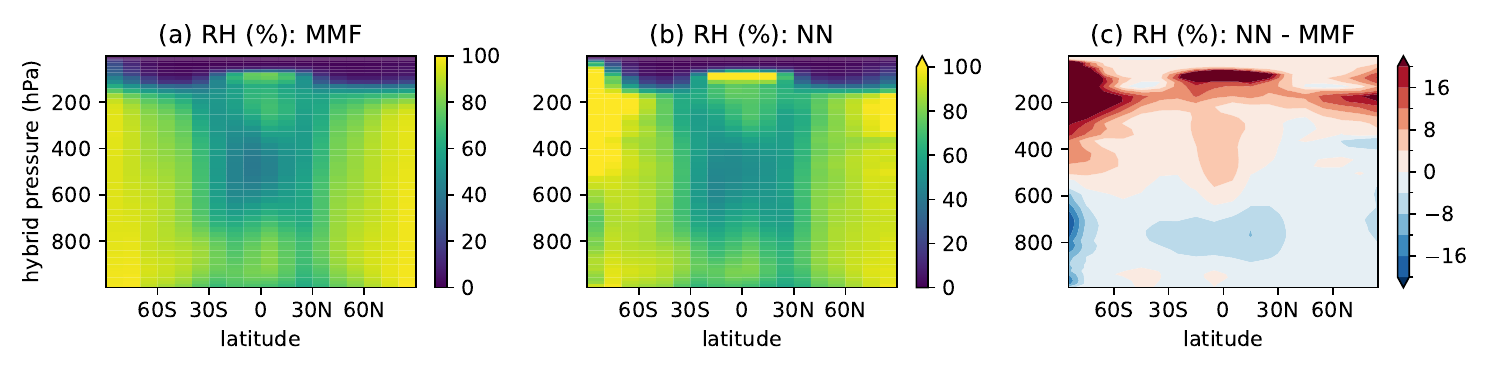}

\caption{Five-year zonal mean relative humidity in the reference E3SM-MMF simulation (left) and in the hybrid simulation with the U-Net-expanded-constrained model (middle). The right column shows the zonal mean bias as the mean state from the hybrid simulation minus that of the reference simulation. The contouring interval in the right column is 4\%.}
\label{fig:rh}
\end{figure}